\documentclass[onecolumn,amsmath,amssymb,nofootinbib,aapt]{revtex4} 

\usepackage{amsmath}
\usepackage{graphicx}% Include figure files
\usepackage{graphics}
\usepackage{dcolumn}% Align table columns on decimal point
\usepackage{bm}% bold math
\usepackage{hyperref}
\usepackage{xcolor}

\def\BEq{\begin{equation}}
\def\EEq{\end{equation}}
\def\BEqA{\begin{eqnarray}}
\def\EEqA{\end{eqnarray}}
\def\BEn{\begin{enumerate}}
\def\EEn{\end{enumerate}}
\def\BWT{\begin{widetext}}
\def\EWT{\end{widetext}}

\begin{document}

\title{Logarithmically enhanced hyperbolic square-root deformation 
of Starobinsky inflation}

\author{Andrei Galiautdinov}
\email{ag1@uga.edu}
\affiliation{
Department of Physics and Astronomy, 
University of Georgia, Athens, Georgia 30602, USA
}

\begin{abstract}
We propose and analyze an enhanced hyperbolic square-root (HSQRT) 
deformation of the Starobinsky model in the context of $f(R)$ gravity. 
The original HSQRT construction provided a globally regular modification 
of $R^2$ inflation, curing the strong-coupling singularity at negative 
curvatures while preserving the characteristic exponential slow-roll 
Starobinsky plateau at large positive curvatures. Motivated by recent precision 
cosmological data (ACT DR6 and DESI) that indicate an upward 
shift in the scalar spectral index and a preference for deviations from pure 
exponential plateau behavior, we introduce a structurally minimal, 
rational-logarithmic enhancement. This phenomenological 
enhancement modifies the deep ultraviolet asymptotic regime while 
preserving global tachyon-free stability, ghost freedom, and the 
recovery of general relativity at low curvatures. In the Einstein frame, the scalaron dynamics 
is described by a globally defined, Pad\'{e}-regulated effective potential, 
$V_{\text{Pad\'{e}}}(\phi) 
= 
V_{\text{Staro}}(\phi) \left(1+ \frac{4\beta }{\frac{2}{3}\kappa^2\phi^2 + 1}\right)^{-1} 
+ \frac{1}{4\alpha\kappa^2} \left[1 - \left( 1 + \sqrt{\frac{2}{3}}\kappa\phi\right)
e^{-\sqrt{\frac{2}{3}}\kappa\phi} \right] e^{-\sqrt{\frac{2}{3}}\kappa\phi}$,
which exhibits the inverse-power asymptotic form at high curvatures, 
$V(\phi) \simeq \frac{1}{8\alpha\kappa^2} \left( 1 - \frac{6\beta}{\kappa^2\phi^2} \right)$, 
with the strength of the deviation from the baseline HSQRT geometry governed 
by a single dimensionless coupling parameter $\beta$. We derive analytic 
expansions for the slow-roll observables, yielding a parameter-free 
leading-order spectral index $n_s \simeq 1 - 3/(2N)$ and a tunable 
tensor-to-scalar ratio $r \simeq 2\sqrt{3\beta}/N^{3/2}$. For standard 
inflationary e-fold durations ($N \in [50, 60]$), this model drives the spectral 
index directly into the newly favored observational window 
($n_s \simeq 0.970\text{--}0.975$) and predicts an exceptionally small 
spectral running $\alpha_s \simeq -3/(2N^2) \in [-0.00060, -0.00042]$, establishing a 
viable, parameter-controlled target for next-generation cosmic microwave 
background observatories.
\end{abstract}

\maketitle

\section{Introduction}

Inflationary cosmology \cite{Guth1981, Linde1982, AlbrechtSteinhardt1982, 
Linde1983, Linde:1984ir, Hawking1982, GuthPi1982, BardeenSteinhardtTurner1983, 
Mukhanov1985, Sasaki1986, LythRiotto1999, BassettTsujikawaWands2006, 
Baumann2009, GuthKaiserNomura2014, MukhanovBook2005, Weinberg2008} 
remains one of the most compelling theoretical frameworks for explaining the observed 
large-scale homogeneity, isotropy, and near scale-invariance of primordial 
fluctuations. Among the wide class of inflationary models, Starobinsky 
inflation \cite{Starobinsky1980, MukhanovChibisov1981, Starobinsky1982, 
Starobinsky1983, Vilenkin1985, Linde2025}, based on quadratic curvature corrections 
in $f(R)$ gravity \cite{Capozziello2008, Capozziello2011, Nojiri2007, Nojiri2011, 
Nojiri2017, DeFelice2010, Sotiriou2010}, continues to provide a robust fit 
to baseline observational data \cite{PlanckInflation2018, BICEPKeck2021} 
without the need to introduce fundamental scalar fields by hand. 

However, recent combinations of high-resolution cosmic microwave 
background (CMB) and baryon acoustic oscillation (BAO) data suggest 
an emerging preference for a slightly higher scalar spectral index, 
$n_s \gtrsim 0.97$ 
\cite{ACT:2025, ACT:2025-2, DESI:2024, DESI:2025, 
Ferreira2025, McDonough2025}. 
This upward shift indicates that explicit deviations from the pure 
exponential plateau structure, which is strictly characteristic of standard 
Starobinsky-type potentials, may be required to maintain precise 
phenomenological viability in the next decade of precision cosmology. 
Consequently, this observational shift has sparked a significant surge 
of theoretical interest, leading to a variety of proposals aimed at modifying 
or extending the standard inflationary attractor framework. While some 
authors appropriately urge caution pending a definitive resolution of the 
underlying CMB-BAO data tension \cite{Ferreira2025}, a wide array of 
phenomenological deformations has rapidly emerged to explicitly 
accommodate this heightened spectral tilt (see, e.\ g., \cite{Ferreira2025} 
and references therein for a comprehensive overview of recent models).

As a foundation for addressing these constraints, the recently proposed 
hyperbolic square-root (HSQRT) deformation \cite{HSQRT} introduces 
a globally regular modification of the Starobinsky model. The defining 
feature of this baseline construction is the replacement of the linear 
derivative,
\BEq
\label{eq:f'Staro}
f_{\text{Staro}}'(R) = 1 + 2\alpha R,
\EEq 
of the $\Lambda$-Starobinsky Lagrangian,
\BEq
\label{eq:fStaro}
f_{\text{Staro}}(R) = R + \alpha R^2 - 2 \Lambda,
\EEq 
used in the the gravitational 
action, 
\BEq
S = \frac{1}{2\kappa^2}\int d^4x \sqrt{-g}\, f(R), \quad 
\kappa^2 = 8\pi G, 
\EEq
with a globally positive, hyperbolic square-root expression,
\begin{equation}
\label{eq:f'Rbase}
f_{\text{base}}'(R) = \alpha R + \sqrt{\alpha^2 R^2 + 1} > 0,
\end{equation}
which integrates to\footnote{Note that one may alternatively 
write, $\mathrm{arcsinh}(\alpha R) 
= \ln (\alpha R + \sqrt{\alpha^2 R^2 + 1})$.}
\begin{equation}
\label{eq:fRbase-arcsinh} 
f_{\text{base}}(R) = \frac{1}{2}R \left(\alpha R + \sqrt{\alpha^2 R^2 + 1}\right) 
+ \frac{1}{2\alpha} \mathrm{arcsinh}(\alpha R) - 2 \Lambda ,
\end{equation}
resulting, upon the introduction of the canonical scalaron field,
\BEq
\kappa \phi = \sqrt{\frac{3}{2}} \ln f_{\text{base}}'(R),
\EEq
in the exact Einstein-frame potential,
\begin{equation}
\label{eq:Vexact}
V_{\text{base}}(\phi) 
= 
\frac{1}{8\alpha\kappa^2} 
\left[ 1 - \left(1 + 2\sqrt{\frac{2}{3}}\kappa\phi\right) 
\exp\left(-2\sqrt{\frac{2}{3}}\kappa\phi\right) \right] 
+ \frac{\Lambda}{\kappa^2} \exp\left(-2\sqrt{\frac{2}{3}}\kappa\phi\right).
\end{equation} 
This geometric structure ensures analytic behavior across all 
scalar curvatures, $R \in (-\infty, +\infty)$, resolving the 
strong-coupling singularity that plagues standard $R^2$ gravity 
when $f'(R) \to 0$ at finite $R$. A critical physical consequence of this geometry 
is the behavior of the scalaron mass, given by
\begin{equation}
\label{eq:exact_mass}
m_s^2(R) 
= \frac{1}{3}\left(\frac{f_{\text{base}}'(R)}{f_{\text{base}}''(R)} - R\right) 
= \frac{1}{3 \alpha} \left( \sqrt{\alpha^2 R^2 + 1} - \alpha R \right)
= \frac{1}{3\alpha f_{\text{base}}'(R)} > 0.
\end{equation}
Because the effective gravitational coupling in scalar-tensor mappings 
is defined as $G_{\text{eff}} = G / f'(R)$, this algebraic HSQRT structure uniquely 
binds the mass of the scalaron directly to the 
running of gravity itself, ensuring globally tachyon-free dynamics. This 
strict inverse proportionality results in the following dynamical behavior 
of the scalaron across the three critical cosmological regimes:
\begin{enumerate}
\item \textbf{Low Curvature ($R \to 0$):} As the universe relaxes to 
the present-day vacuum, $f_{\text{base}}'(R) \to 1$. The mass smoothly approaches 
$m_s^2 \to 1/(3\alpha)$, safely evading local macroscopic fifth-force 
constraints.
\item \textbf{Deep Inflation ($R \to +\infty$):} In the early universe, 
$f_{\text{base}}'(R) \simeq 2\alpha R \to +\infty$, causing the scalaron to become 
dynamically ultra-light ($m_s^2 \to 0$), guaranteeing an exceptionally 
flat energetic potential for slow-roll.
\item \textbf{Negative Curvature Boundary ($R \to -\infty$):} As 
spacetime approaches the strong-coupling boundary, $f_{\text{base}}'(R) \to 0^+$. 
The scalar degree of freedom becomes infinitely heavy 
($m_s^2 \to +\infty$) and dynamically freezes out, fortifying a divergent 
energetic barrier that preventing pathological crossing into the ghost regime.
\end{enumerate}

While the baseline HSQRT model successfully cures the classical 
singularities of $R^2$ gravity, its unperturbed geometry acts as 
a rigorous universal attractor \cite{Kallosh2013AlphaAttractors, 
Kallosh2013AlphaAttractorsReview}. It yields the standard Starobinsky 
scalar spectral index $n_s \simeq 1 - 2/N \simeq 0.966$ and a strongly 
suppressed tensor-to-scalar ratio of $r \simeq 3/N^2 \simeq 0.00083$. 
While safe from current upper bounds, this $n_s$ prediction falls 
slightly below the newly favored observational window. To systematically 
accommodate the 2024--2026 data shifts, the asymptotic behavior of 
the model at large curvatures must be modified to break the conformal 
attractor trap.

Ideally, such modifications should be physically motivated by the 
high-energy expectations of the gravitational effective action 
\cite{QuantumGravity1, QuantumGravity2, Barvinsky2023a, 
Barvinsky2023, Cognola2005, Mirzabekian1996}. In this paper, we propose 
an enhanced logarithmic deformation of the baseline HSQRT model. 
Motivated by functional renormalization-group (FRG) flows 
\cite{Bonanno2002RGInflation, Reuter2012AsymptoticSafety, 
Alexandre2025RGFlowACT} and the structure of Pad\'{e}-type 
resummations \cite{Baker1996}, we introduce a mathematically 
exact, rational-logarithmic ansatz governed by a single dimensionless 
ultraviolet coupling parameter $\beta$. In doing so, we utilize 
the globally positive HSQRT geometry as an indispensable 
\emph{algebraic regularizer} to protect the logarithmic enhancement
from generating imaginary branch cuts at negative curvatures.

Because the inclusion of logarithmic terms generically prevents 
the exact algebraic inversion of the conformal mapping, we develop 
an exact parametric formulation of the Einstein-frame dynamics 
governed by a purely geometric derivative variable $y$
(Sec.\ \ref{sec:ParametricConstruction}). This analytical methodology 
allows us to shift the ultraviolet approach of the scalar potential from 
an exponential to an inverse-power scaling ($1/\phi^2$) without 
relying on piecewise approximations. As we will demonstrate below, 
this phenomenologically motivated extension asymptotically vanishes in the 
infrared regime to preserve the rigorous global stability of the baseline theory. 
Simultaneously, in the deep ultraviolet regime, it drives the scalar 
spectral index to the observationally favored value of $n_s \simeq 0.975$ 
at e-fold number $N = 60$ while yielding a tunable, highly 
testable tensor-to-scalar ratio for next-generation observatories.

Because the logical progression underlying the derivation of various 
deformation forms of Starobinsky theory is somewhat elaborate, 
for the reader's convenience, 
we include Table \ref{tab:1} summarizing the primary equations of the paper. 
This table traces the conceptual sequence from the standard Starobinsky model 
to the baseline HSQRT geometry (presented in standard and alternative algebraic forms), 
through the linearly enhanced HSQRT model, and finally to the hypothesized 
Pad\'{e}-regularized effective potential which robustly generalizes the 
enhanced proposal.
%%%%%%%%%%% BEGIN TABLE 1:
\begin{table}%[!ht]
\caption{\label{tab:1} Summary of the main equations used 
to define various deformation forms of the Starobinsky model 
considered in this article.}
\begin{ruledtabular}
\begin{tabular}{lcccc}
     & {\bf Starobinsky }& {\bf base HSQRT}& {\bf enhanced HSQRT}& 
{\bf Pad\'{e}-regularized enhanced HSQRT}
\\
\hline
\\
{\bf Notation:} &
$f_{\text{Staro}}(R)$, $V_{\text{Staro}}(\phi)$ & 
$f_{\text{base}}(R)$, $V_{\text{base}}(\phi)$ & 
$f(R)$, $V(\phi)$  & 
$f_{\text{Pad\'{e}}}(R)$, $V_{\text{Pad\'{e}}}(\phi)$
\\
\\
{\bf Standard form:} &
(\ref{eq:f'Staro}), (\ref{eq:fStaro}), (\ref{eq:VexactStarobinsky}) & 
(\ref{eq:f'Rbase}), (\ref{eq:fRbase-arcsinh}), 
(\ref{eq:Vexact}), (\ref{eq:f'base}), (\ref{eq:fybase}) & 
(\ref{eq:explicit_log_model-standard}), (\ref{eq:explicit_log_model_R}),
(\ref{eq:V_global_beta})  & 
\\
\\
{\bf Alternative form:} &
 & (\ref{eq:fybase-alt}) & (\ref{eq:explicit_log_model}), (\ref{eq:V_global_beta_alt})  & 
(\ref{eq:V_global_beta_true}),  (\ref{eq:RPade_parametric}), (\ref{eq:fyPade_parametric}) 
\\
\\
\end{tabular}
\end{ruledtabular}
\end{table}
%%%%%%%%%%% END TABLE 1

\section{The Enhanced HSQRT Model}

In conformal mappings of $f(R)$ gravity, the $n_s\simeq 1-2/N$ 
prediction is an asymptotic trap inherent to any potential that 
approaches the inflationary plateau exponentially, 
\BEq
V(\phi)\simeq V_0 \left(1-{\cal A}(\phi)e^{-\lambda \phi}\right),
\EEq
where ${\cal A}(\phi)$ varies much slower than the exponential. 
This exponential behavior is guaranteed whenever the derivative 
of the Lagrangian is dominated by a linear term, $f'(R) \propto R$, 
at high curvatures. To raise the spectral index and accommodate the latest 
precision data \cite{Ferreira2025, Kallosh2025Status}, the Einstein-frame 
potential must transition to an inverse-power plateau 
(see Eq.\ (6.318), with $p=2$, in Sec.\ 6.19, ``Brane Inflation (BI)'' 
of Ref.\ \cite{Martin2014} (arXiv v5, 2024); 
cf.\ \cite{Roest2014, GarciaBellido2014}),
\BEq
V(\phi) \simeq V_0 (1 - \mu / \phi^2).
\EEq
This specific $1/\phi^2$ asymptotic universality class has recently 
been shown by Kallosh and Linde to be a plausible 
phenomenological possibility to address the $n_s$ tension, 
as this potential shape naturally emerges in supergravity 
setting as polynomial \cite{Kallosh2022Polynomial} or singular 
\cite{Kallosh2025Singular} $\alpha$-attractors. 
Similarly, extensions of the Starobinsky model 
motivated by the ACT results have recently been realized within 
the framework of no-scale supergravity \cite{Gialamas2025ACTive}. 
Furthermore, the exact same asymptotic scaling is successfully 
generated in scalar-tensor theories by the simplest chaotic 
inflation models endowed with a linear non-minimal coupling 
to gravity, $(1+\phi)R$ \cite{Kallosh2025ACT}. 

It is then very natural to ask if the same phenomenological 
target can be reached through purely geometric degrees of freedom 
(cf., e.\ g., \cite{Gialamas2026Weyl}). Since in $f(R)$ gravity, the conformal 
scalar field scales logarithmically with the curvature derivative, 
$\kappa \phi \propto \ln f'(R)$, achieving the favorable inverse-power 
plateau requires introducing a logarithmic running into the 
high-curvature scaling of the spacetime geometry. 

\begin{figure}[t]%[!ht]
\centering
\includegraphics[angle=0,width=0.55\linewidth]{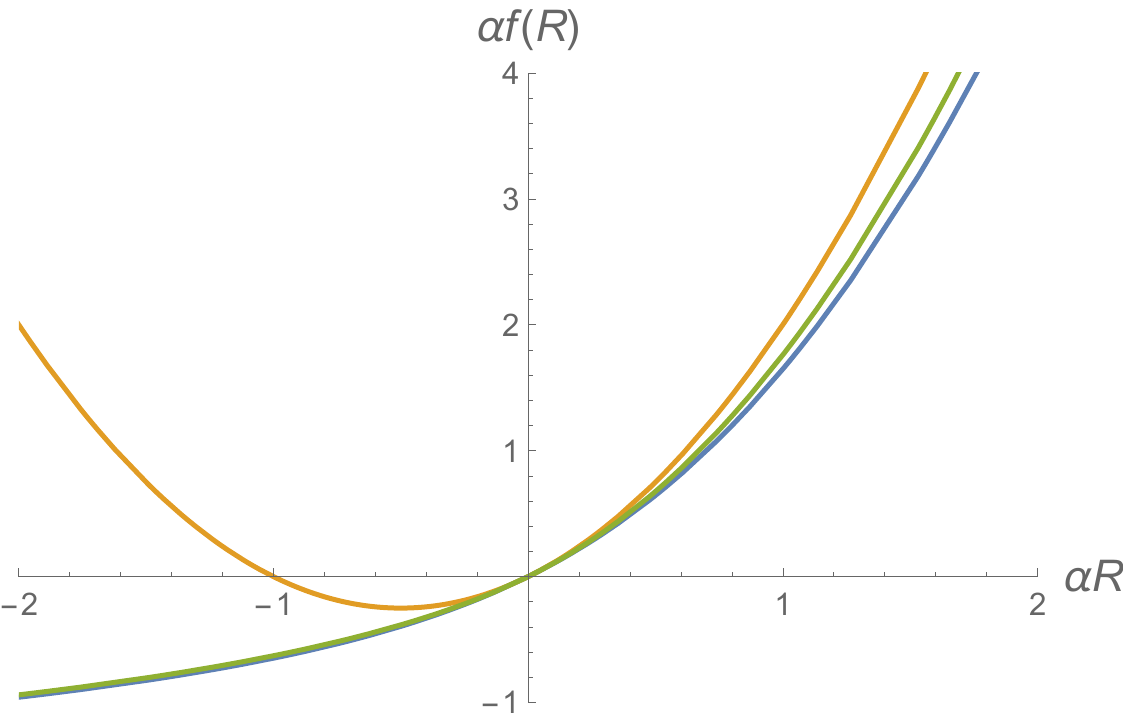}
\caption{\label{fig:01} The Lagrangian densities, $f(R)$, of the 
base HSQRT (blue; Eq.\ (\ref{eq:fRbase-arcsinh})), 
enhanced HSQRT (green; Eq.\ (\ref{eq:explicit_log_model_R})), 
and standard Starobinsky (orange; Eq.\ (\ref{eq:fStaro})) models, 
at $\beta = 0.1$ with $\Lambda=0$. Both the base and enhanced 
HSQRT functions recover the linear Einstein-Hilbert limit,
$f(R) \simeq R$, at low curvatures, while transitioning to, respectively, 
the standard Starobinsky, $f(R) \simeq  \alpha R^2$, and improved Starobinsky,
$f(R) \simeq \alpha R^2 \left[ 1 + \frac{4\beta}{\ln^2(2\alpha R)} \right]$, 
scalings required for slow-roll inflation at large positive curvatures.}
\end{figure}

\begin{figure}[t]%[!ht]
\centering
\includegraphics[angle=0,width=0.55\linewidth]{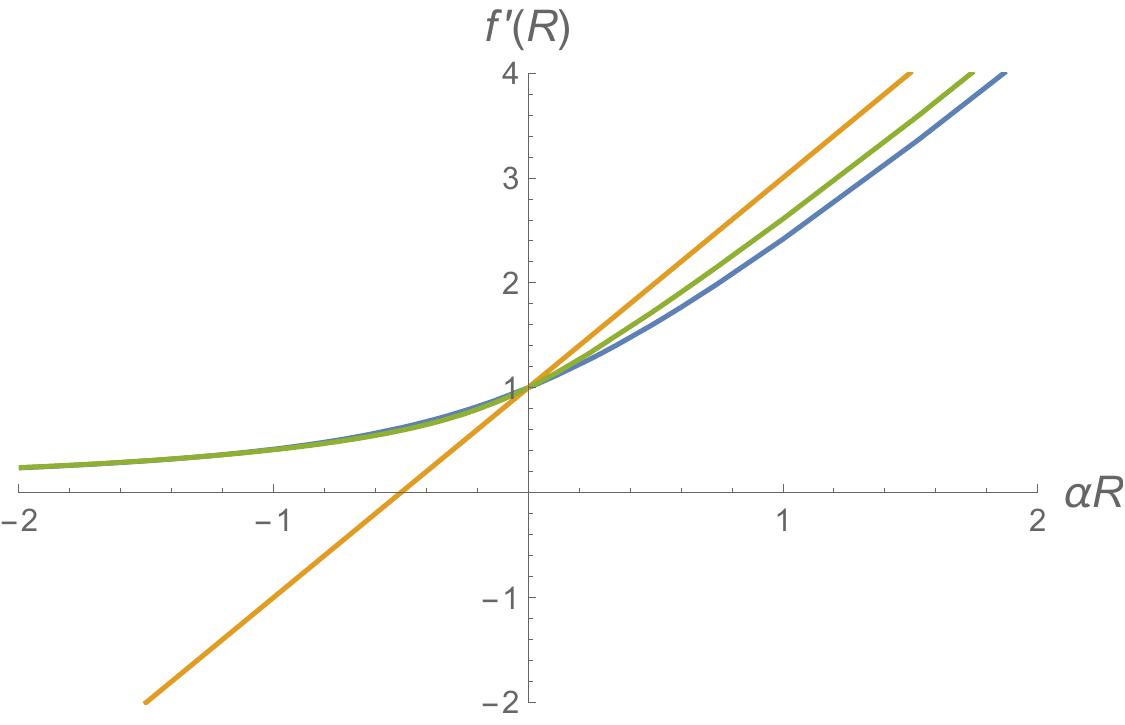}
\caption{\label{fig:02} The first derivatives of the Lagrangian densities, 
$f'(R)$, in the 
base HSQRT (blue; Eq.\ (\ref{eq:fRbase-arcsinh})), 
enhanced HSQRT (green; Eq.\ (\ref{eq:explicit_log_model_R})), 
and standard Starobinsky (orange; Eq.\ (\ref{eq:fStaro})) models, 
at $\beta = 0.1$. In both the base and enhanced 
HSQRT models, the function $f'(R)$ remains strictly positive 
for all real curvature values $R \in (-\infty, +\infty)$. 
This positivity ensures that the effective gravitational 
coupling $G_{\mathrm{eff}} = G/f'(R)$ is finite and positive, 
preventing the spin-2 graviton from becoming a ghost and ensuring 
the conformal mapping to the Einstein frame remains non-singular.}
\end{figure}

\begin{figure}[!ht]
\centering
\includegraphics[angle=0,width=0.55\linewidth]{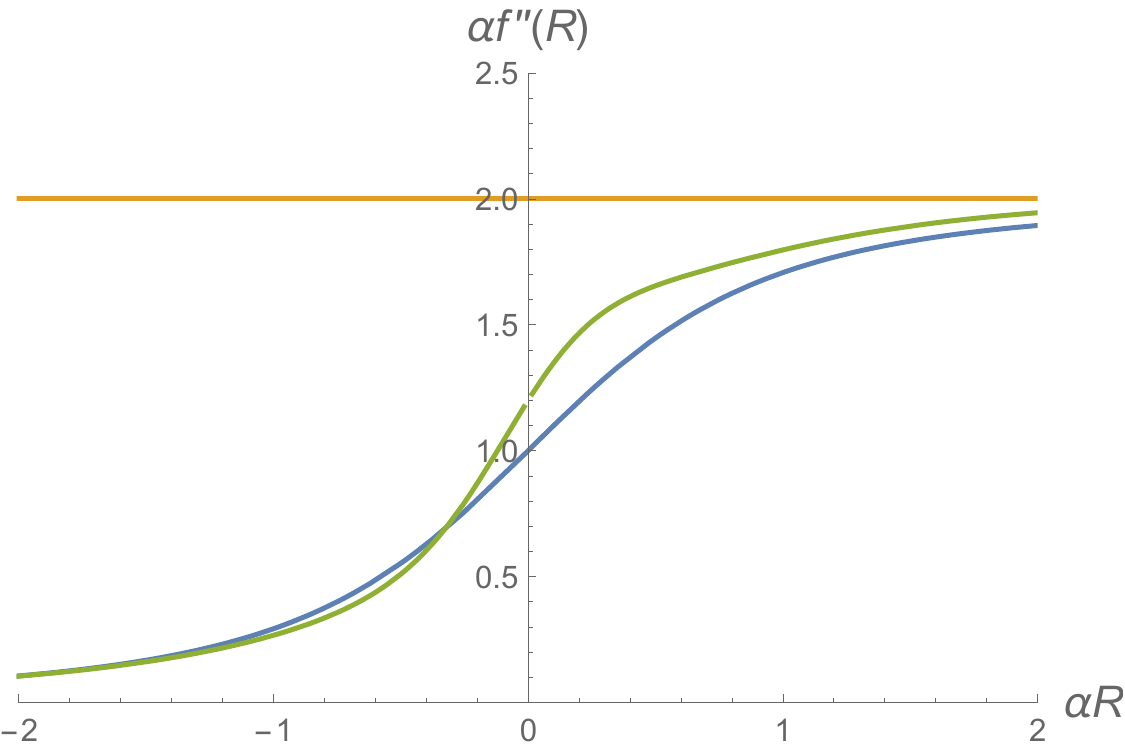}
\caption{\label{fig:03} The second derivatives of the Lagrangian densities, 
$f''(R)$, in the 
base HSQRT (blue; Eq.\ (\ref{eq:fRbase-arcsinh})), 
enhanced HSQRT (green; Eq.\ (\ref{eq:explicit_log_model_R})), 
and standard Starobinsky (orange; Eq.\ (\ref{eq:fStaro})) models, 
at $\beta = 0.1$. The strict positivity $f''(R) > 0$ in both the base 
and enhanced HSQRT models is maintained across the 
entire curvature domain $R \in (-\infty, +\infty)$. 
This is a necessary condition in $f(R)$ gravity to avoid the 
Dolgov-Kawasaki instability and ensures the conformal mapping to 
the Einstein frame remains non-singular.}
\end{figure}

\begin{figure}[!ht]
\centering
\includegraphics[angle=0,width=0.55\linewidth]{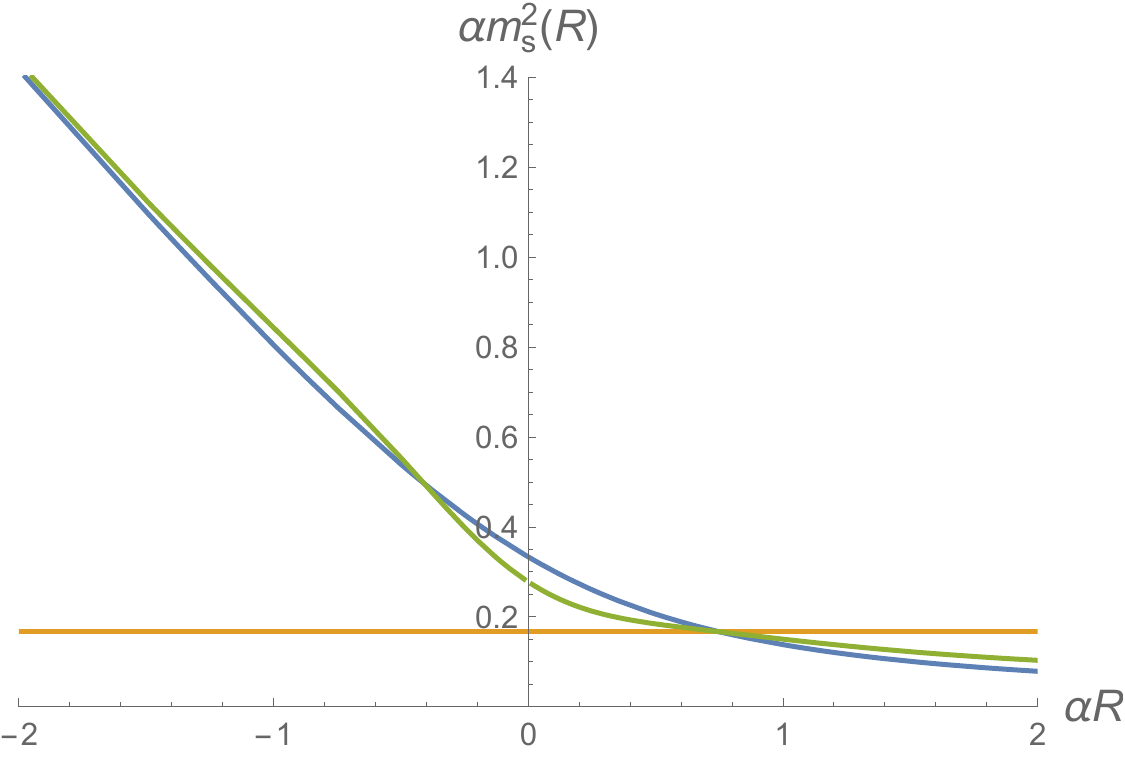}
\caption{\label{fig:04} The effective scalaron mass squared, 
$m_{\mathrm{s}}^2$, in the 
base HSQRT (blue; Eq.\ (\ref{eq:fRbase-arcsinh})), 
enhanced HSQRT (green; Eq.\ (\ref{eq:explicit_log_model_R})), 
and standard Starobinsky (orange; Eq.\ (\ref{eq:fStaro})) models, 
at $\beta = 0.1$. Note that $m_{\mathrm{s}}^2$ remains strictly 
positive across the entire global curvature domain 
$R \in (-\infty, +\infty)$, ensuring the classical and 
perturbative stability of the spacetime, preventing 
the scalar degree of freedom from becoming tachyonic.}
\end{figure}

Correspondingly, and rather than defining the modified theory 
through a complicated non-integrable derivative, we construct 
an exact, closed-form Lagrangian by exploiting the algebraic 
simplicity of the base hyperbolic square-root model. 
Let us write the baseline derivative and baseline Lagrangian 
in the form, 
\BEq
\label{eq:f'base}
y(R) \equiv f_{\text{base}}'(R) = \alpha R + \sqrt{\alpha^2 R^2 + 1}, 
\EEq
and
\begin{equation}
\label{eq:fybase}
f_{\text{base}}(y) = \frac{y^2 - 1}{4\alpha} + \frac{\ln y}{2\alpha} ,
\end{equation}
respectively, with the integration constant set to zero, $\Lambda=0$, 
for simplicity. To better illuminate the underlying algebraic architecture 
of the baseline theory, we can complete the square to rewrite the unperturbed 
Lagrangian as the difference of two distinct geometric sectors,
\begin{equation}
\label{eq:fybase-alt}
f_{\text{base}}(y) = \frac{y^2 - y + \ln y}{2\alpha} - \frac{(y-1)^2}{4\alpha}.
\end{equation}
To implement the logarithmic running while structurally preserving the 
low-energy limit, we introduce a dimensionless 
ultraviolet coupling parameter $\beta \ll 1$ and define the enhanced 
Lagrangian by modulating the pure $(y-1)^2$ sector with a positive 
rational-logarithmic regulator that ensures that the inflationary potential 
approaches the plateau from below,
\begin{equation}
\label{eq:explicit_log_model}
f(y) 
= \frac{y^2 - y + \ln y}{2\alpha} - \frac{(y-1)^2}{4\alpha} 
\left( 1 - \frac{4\beta}{\ln^2 y + 1} \right).
\end{equation}
Expanding this representation yields the explicitly separated form, 
\BEq
\label{eq:explicit_log_model-standard}
f(y) 
= f_{\text{base}}(y) + f_{\text{corr}}(y)
= \left(\frac{y^2 - 1}{4\alpha} + \frac{\ln y}{2\alpha}\right)
+ \frac{\beta}{\alpha} \frac{(y - 1)^2}{\ln^2 y + 1},
\EEq 
where the correction term is 
\begin{equation}
f_{\text{corr}}(y) = \frac{\beta}{\alpha} \frac{(y - 1)^2}{\ln^2 y + 1},
\end{equation}
producing
\begin{equation}
\label{eq:explicit_log_model_R}
f(R) 
= 
\frac{\left(\alpha R + \sqrt{\alpha^2 R^2 + 1}\right)^2 - 1}{4\alpha} 
+ \frac{\ln \left(\alpha R + \sqrt{\alpha^2 R^2 + 1}\right)}{2\alpha}
+ \frac{\beta}{\alpha} \frac{(\alpha R + \sqrt{\alpha^2 R^2 + 1} - 1)^2}
{\ln^2 \left(\alpha R + \sqrt{\alpha^2 R^2 + 1}\right) + 1},
\end{equation}
as shown in Figs.\ \ref{fig:01}, \ref{fig:02}, \ref{fig:03}, and \ref{fig:04}.
Note that at small curvatures ($|\alpha R|\ll1$) this Lagrangian
behaves as
\BEq
f(R)
=
R
+\left( \frac{1}{2} + \beta \right) \alpha R^2 
+\left(\frac{1}{6} + \beta \right) \alpha^2 R^3 
-\frac{3}{4} \beta \alpha ^3 R^4  
+{\cal O}\left(R^5\right),
\EEq
safely recovering standard general relativity. On the other hand, 
in the deep ultraviolet regime at large curvatures 
($R \to +\infty$), using
\BEq
\alpha R + \sqrt{\alpha^2 R^2 + 1} \simeq 2\alpha R
\EEq 
in Eq.\ (\ref{eq:explicit_log_model_R}), we get: 
for the first term (baseline $R^2$ component),
\begin{equation}
\frac{\left(\alpha R + \sqrt{\alpha^2 R^2 + 1}\right)^2 - 1}{4\alpha} 
\simeq
\frac{4\alpha^2 R^2}{4\alpha} = \alpha R^2,
\end{equation}
reproducing the quadratic scaling of the standard Starobinsky action; 
for the second term (baseline logarithmic sub-component),
\begin{equation}
\frac{\ln \left(\alpha R + \sqrt{\alpha^2 R^2 + 1}\right)}{2\alpha}
\simeq
\frac{\ln(2\alpha R)}{2\alpha},
\end{equation}
which grows merely logarithmically and becomes dynamically 
negligible; for the third term ($\beta$-parametrized 
rational-logarithmic  enhancement), 
\begin{equation}
\frac{\beta}{\alpha} \frac{(\alpha R + \sqrt{\alpha^2 R^2 + 1} - 1)^2}
{\ln^2 \left(\alpha R + \sqrt{\alpha^2 R^2 + 1}\right) + 1}
\simeq
\frac{4 \alpha \beta R^2}{\ln^2(2\alpha R)},
\end{equation}
yielding the dominant high-energy modification, and, thus, 
leading to the complete asymptotic Lagrangian,
\begin{equation}
\label{eq:fR_UV_limit}
f(R) \simeq \alpha R^2 \left[ 1 + \frac{4\beta}{\ln^2(2\alpha R)} \right].
\end{equation}

%%%%%%%%% BEGIN FIG. 1
%%%%%%%%% END FIG. 1

%%%%%%%%% BEGIN FIG. 2
%%%%%%%%% END FIG. 2

%%%%%%%%% BEGIN FIG. 3
%%%%%%%%% END FIG. 3

%%%%%%%%% BEGIN FIG. 4
%%%%%%%%% END FIG. 4

\section{Exact Parametric Construction of the Enhanced Model 
and Global Scalaron Potential}
\label{sec:ParametricConstruction}

In the unperturbed baseline model ($\beta = 0$), the conformal 
mapping to the Einstein frame is governed by the simple relation 
(\ref{eq:f'base}), which allows for a globally exact, closed-form 
inversion to yield $V(\phi)$, shown in Eq.\ (\ref{eq:Vexact}).
However, the introduction of the $\beta$-modulated correction 
explicitly breaks the algebraic invertibility 
of the conformal factor. Consequently, the complete and globally exact 
dynamics of the Einstein frame must be rigorously expressed via 
a parametric formulation governed by $y \in (0, +\infty)$.

We begin with the exact inverse curvature mapping of the baseline 
theory. By algebraically inverting the definition of $y$, the Ricci scalar 
is parameterized as
\begin{equation}
R(y) = \frac{y^2 - 1}{2\alpha y},
\end{equation}
with the associated Jacobian of the transformation being 
\BEq
\frac{dR}{dy} = \frac{y^2 + 1}{2\alpha y^2}.
\EEq
To determine the exact conformal factor, 
\BEq
F(y) \equiv f'(R),
\EEq 
for the modified theory, we define the correction function,
\BEq
H(y) = \frac{(y - 1)^2}{\ln^2 y + 1},
\EEq 
whose first derivative with respect to the parametric variable $y$ is given by
\begin{equation}
\label{eq:H'(y)}
H'(y) 
= 
\frac{2(y - 1)}{\ln^2 y + 1} \left[ 1 - \frac{(y - 1)\ln y}{y(\ln^2 y + 1)} \right].
\end{equation}
Applying the chain rule, $F(y) = \frac{df/dy}{dR/dy}$, and recognizing 
that the baseline component simply yields $y$, the exact conformal 
factor takes the closed form,
\begin{equation}
\label{eq:F(y)}
F(y) = y + \frac{2\beta y^2}{y^2 + 1} H'(y).
\end{equation}

\begin{figure}[!ht]
\centering
\includegraphics[angle=0,width=0.6\linewidth]{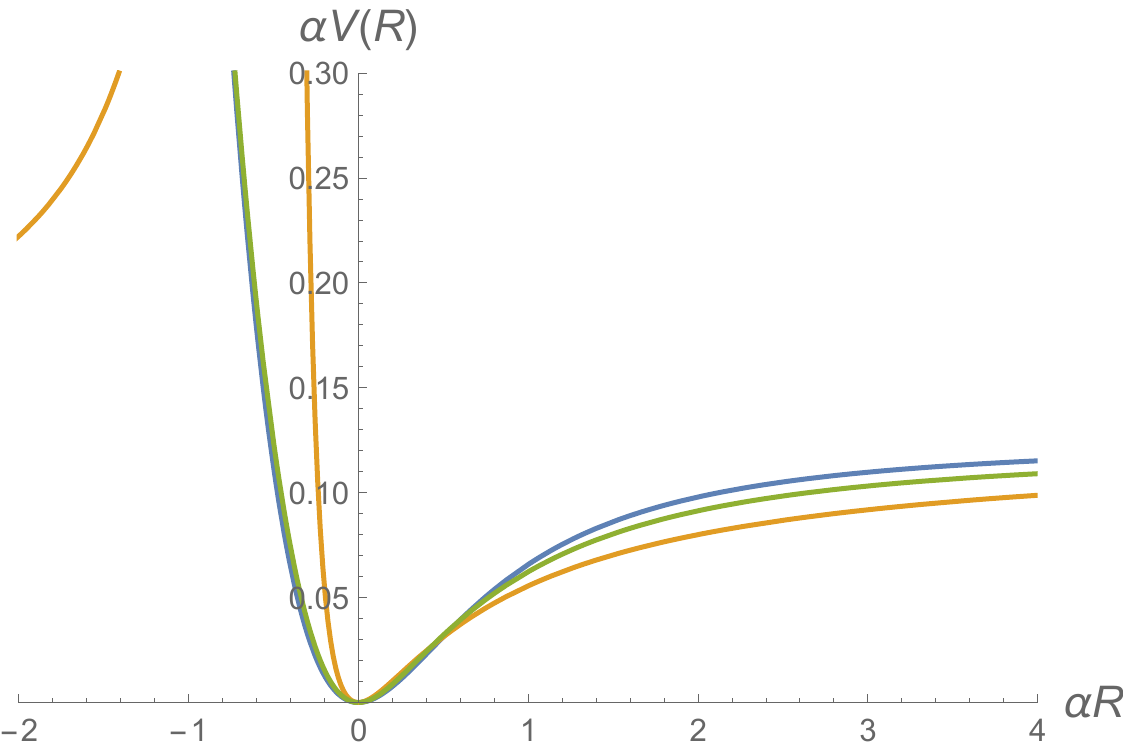}
\caption{\label{fig:05} The scalaron potential expressed as a function 
of the Jordan-frame Ricci scalar, 
$V(R)=(Rf'(R)-f(R))/(2\kappa^2 f'(R)^2)$, $\kappa = 1$, in the 
base HSQRT (blue; Eq.\ (\ref{eq:fRbase-arcsinh})), 
enhanced HSQRT (green; Eq.\ (\ref{eq:explicit_log_model_R})), 
and standard Starobinsky (orange; Eq.\ (\ref{eq:fStaro})) models, 
at $\beta = 0.1$. Note that the standard Starobinsky potential suffers 
a pathological pole at $R_{\mathrm{crit}}=-1/(2\alpha)$, where $f_{\text{Staro}}'(R)=0$. 
At this boundary, the effective gravitational coupling diverges, and the 
conformal mapping becomes singular. In contrast, the deformed models' 
potentials smoothly bypass this singularity, remaining finite and globally 
well-defined for all $R \in (-\infty, +\infty)$.}
\end{figure}

With the conformal factor globally defined, the exact canonical 
scalar field $\phi(y)$ and its corresponding potential $V(y)$ 
(see Fig.\ \ref{fig:05}) are given by the standard conformal 
definitions, parameterized by $y$,
\begin{equation}
\label{eq:parametric_einstein}
\kappa \phi(y) = \sqrt{\frac{3}{2}} \ln F(y),
\quad
V(y) = \frac{R(y)F(y) - f(y)}{2\kappa^2 F(y)^2}.
\end{equation}
This parametric dual-equation system is exact, real, and continuous 
across the entire physical domain $y \in (0, +\infty)$.  

%%%%%%%%% BEGIN FIG. 5
%%%%%%%%% END FIG. 5

\begin{figure}[t]%[!ht]
\centering
\includegraphics[angle=0,width=0.6\linewidth]{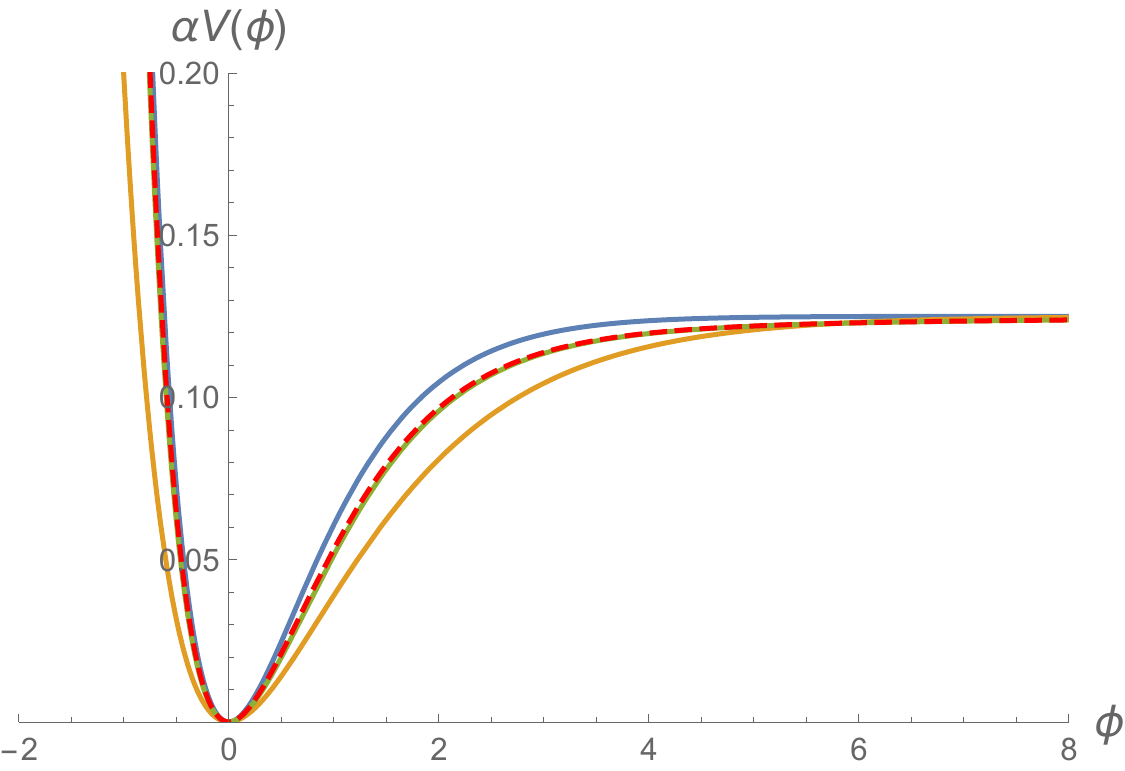}
\caption{\label{fig:06} Einstein-frame scalaron potential, $V(\phi)$, 
in the base HSQRT (blue; Eq.\ (\ref{eq:Vexact})), 
enhanced HSQRT (green; Eq.\ (\ref{eq:V_global_beta})), 
enhanced HSQRT Pad\'{e} approximant 
(dashed red; Eq.\ (\ref{eq:V_global_beta_true})), 
and standard Starobinsky (orange; Eq.\ (\ref{eq:VexactStarobinsky})) models, 
with $\Lambda = 0$ and $\beta = 0.1$.
At large 
positive $\phi$, all potentials asymptote to the inflationary plateau at 
$1/(8\alpha)$, while creating a steep energetic barrier at $\phi \to -\infty$. 
The physical distinction lies in the Jordan-frame 
origin of these walls: the standard Starobinsky wall corresponds to a singular 
truncation of the manifold at a finite curvature $R = -1/(2\alpha)$, beyond 
which the mapping yields unphysical fields. In contrast, the deformed models' 
walls correspond to the true geometric infinity $R \to -\infty$, ensuring the 
global curvature domain $R \in (-\infty, +\infty)$ is non-singularly mapped 
onto the canonical scalar field space.}
\end{figure}

\begin{figure}[!ht]
\centering
\includegraphics[angle=0,width=0.6\linewidth]{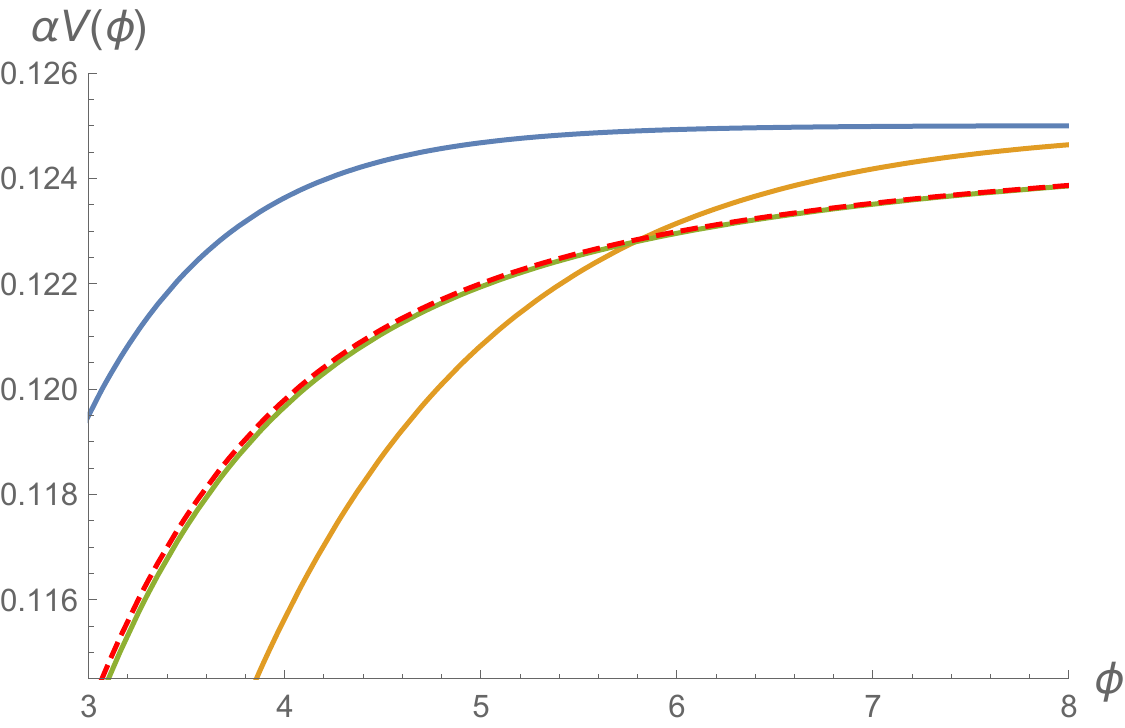}
\caption{\label{fig:07} Asymptotic behavior at large curvatures 
of the Einstein-frame scalaron potential, $V(\phi)$, 
in the base HSQRT (blue; Eq.\ (\ref{eq:Vexact})), 
enhanced HSQRT (green; Eq.\ (\ref{eq:V_global_beta})), 
enhanced HSQRT Pad\'{e} approximant 
(dashed red; Eq.\ (\ref{eq:V_global_beta_true})), 
and standard Starobinsky (orange; Eq.\ (\ref{eq:VexactStarobinsky})) 
models, with $\Lambda = 0$ and $\beta = 0.1$.}
\end{figure}

\begin{figure}[!ht]
\centering
\includegraphics[angle=0,width=0.6\linewidth]{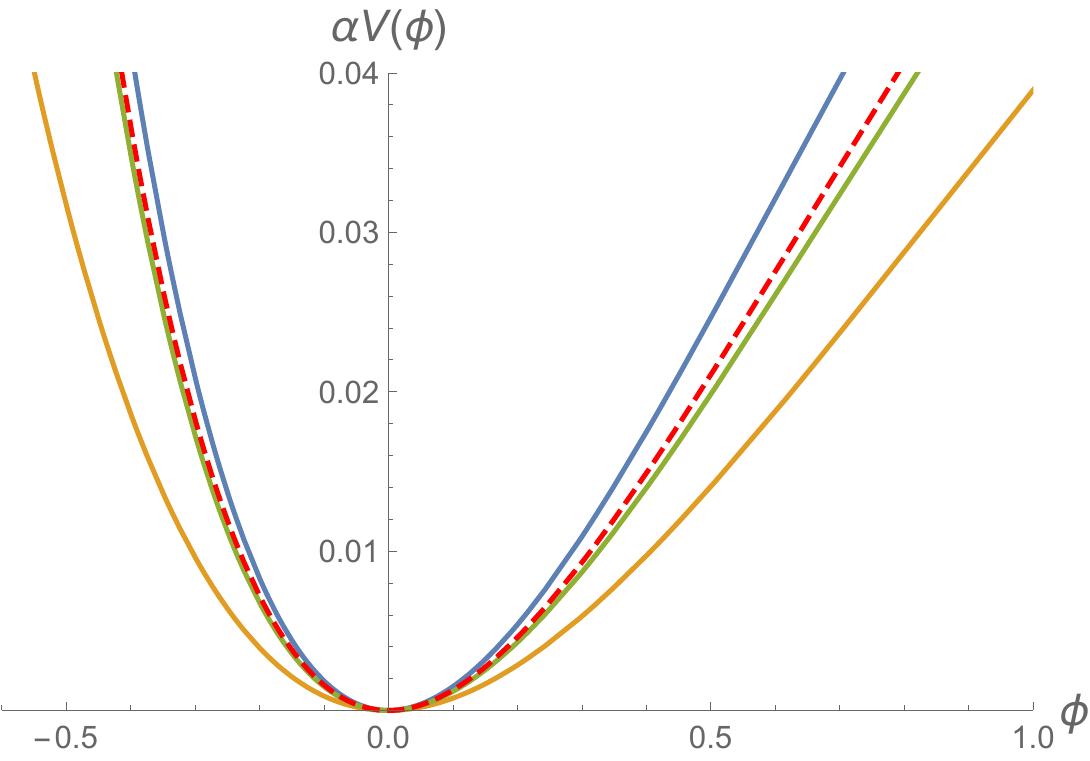}
\caption{\label{fig:08} Asymptotic behavior at small curvatures 
of the Einstein-frame scalaron potential, $V(\phi)$, 
in the base HSQRT (blue; Eq.\ (\ref{eq:Vexact})), 
enhanced HSQRT (green; Eq.\ (\ref{eq:V_global_beta})), 
enhanced HSQRT Pad\'{e} approximant 
(dashed red; Eq.\ (\ref{eq:V_global_beta_true})), 
and standard Starobinsky (orange; Eq.\ (\ref{eq:VexactStarobinsky})) 
models, with $\Lambda = 0$ and $\beta = 0.1$.}
\end{figure}

While an explicit algebraic inversion of the fully non-linear system 
into a single analytic function $V(\phi)$ is transcendentally forbidden, 
the global $\mathcal{O}(\beta)$ approximation of the Einstein-frame 
potential can be derived exactly by exploiting the properties of 
the Legendre transformation. To simplify the discussion,
let us define the rescaled canonical field, 
\BEq
\varphi \equiv \sqrt{2/3}\kappa\phi,
\EEq
with the conformal factor in the Einstein frame being 
\BEq
F(R) = e^{\varphi}.
\EEq 
For a linearly perturbed Lagrangian, 
\BEq
f(R) = f_{\text{base}}(R) + \beta \delta f(R),
\EEq
with the derivative,
\BEq
F(R) = y + {\cal O}(\beta),
\EEq
the exact potential 
$V = (RF - f)/(2\kappa^2 F^2)$ behaves under variations at a fixed 
conformal factor $F$ (and thus a fixed physical field $\varphi$) 
according to the envelope theorem,\footnote{To prove this explicitly, 
let $R_{\text{base}}$ be the curvature 
in the unperturbed theory that yields a specific conformal factor $F$, 
such that $F = f_{\text{base}}'(R_{\text{base}})$. When we linearly perturb the 
Lagrangian, $f = f_{\text{base}} + \beta \delta f$, maintaining this exact same 
fixed $F$ (and thus the same physical point $\varphi$ in the Einstein-frame 
field space) requires the curvature to implicitly shift by some unknown $\delta R$, 
becoming $R = R_{\text{base}} + \delta R$. Expanding the variation of the 
general Einstein-frame potential $V = (RF - f)/(2\kappa^2 F^2)$ to first order 
yields $\delta V = [(\delta R)F - (f_{\text{base}}'(R_{\text{base}})\delta R 
+ \beta \delta f(R_{\text{base}}))]/(2\kappa^2 F^2)$. Because the background 
mapping strictly enforces $F = f_{\text{base}}'(R_{\text{base}})$, 
the $\delta R$ cross-terms cancel, leaving 
$\delta V|_F = -\beta \delta f(R_{\text{base}}) / (2\kappa^2 F^2)$. This result 
demonstrates that the first-order shift in the scalar potential does 
not require the transcendental inversion of the perturbed curvature $R(F)$.}
\begin{equation}
\delta V\Big|_F = - \frac{\beta \delta f|_R}{2\kappa^2 F^2}.
\end{equation}
This identity ensures that any $\mathcal{O}(\beta)$ 
variation in the parametric relationship $R(F)$ strictly cancels out 
in the numerator. Consequently, although the canonical field $\varphi$ 
of the enhanced model differs from that of the base model by a 
$\beta$-dependent correction in the Jordan frame, the first-order 
shift to the scalar potential expressed as a function of this 
exact $\varphi$ is simply the explicit perturbation of the Lagrangian 
evaluated by taking the zeroth-order background limit, 
\BEq
y = F_{\text{base}} \simeq F = e^\varphi,
\EEq
allowing us to substitute $y$ for $e^\varphi$ inside the correction term,
\begin{equation}
V(\varphi) 
\simeq 
V_{\text{base}}(\varphi) - \frac{\beta}{\alpha} \frac{H(e^\varphi)}{2\kappa^2 e^{2\varphi}}.
\end{equation}
Inserting $R(y) = (y - y^{-1})/(2\alpha)$ and $f_{\text{base}}(y)$ 
evaluated at $y = e^\varphi$ recovers the globally 
exact HSQRT baseline potential,
\begin{equation}
V_{\text{base}}(\varphi) 
=
 \frac{Ry-f_{\text{base}}}{2\kappa^2y^2}
=
\frac{1}{2\kappa^2 e^{2\varphi}}
\left[
\frac{e^{2\varphi}-1}{2\alpha}-\frac{e^{2\varphi}-1}{4\alpha}-\frac{\varphi}{2\alpha}
\right]
=
\frac{1}{8\alpha\kappa^2} \left[ 1 - (1+ 2\varphi) e^{-2\varphi} \right].
\end{equation}
Using our explicitly defined rational-logarithmic correction 
$H(e^\varphi) = (e^\varphi - 1)^2 / (\varphi^2 + 1)$, the fractional 
term simplifies as
\begin{equation}
\delta V(\varphi) 
= 
- \frac{\beta}{2\alpha\kappa^2} \frac{(e^\varphi - 1)^2}{e^{2\varphi} (\varphi^2 + 1)} 
= 
- \frac{4\beta}{8\alpha\kappa^2} \frac{(1 - e^{-\varphi})^2}{\varphi^2 + 1}.
\end{equation}
Defining the standard inflationary scale 
\BEq
V_0 \equiv 1/(8\alpha\kappa^2),
\EEq 
we arrive at the exact, closed-form global scalar potential expanded 
to linear order in $\beta$ (Fig.\ \ref{fig:06}),
\begin{align}
\label{eq:V_global_beta}
V(\phi) 
&=
V_0 \left[ 
1 - \left( 1 + 2\sqrt{\frac{2}{3}}\kappa\phi\right) 
e^{-2\sqrt{\frac{2}{3}}\kappa\phi}
- \frac{4\beta \left(1 - e^{-\sqrt{\frac{2}{3}}\kappa\phi}\right)^2}
{\frac{2}{3}\kappa^2\phi^2 + 1} \right]
\nonumber \\
&\equiv
V_{\text{base}}(\phi) - \frac{4\beta V_0 \left(1 - e^{-\sqrt{\frac{2}{3}}\kappa\phi}\right)^2}
{\frac{2}{3}\kappa^2\phi^2 + 1}.
\end{align}
%%%%%%%%% BEGIN FIG. 6
%%%%%%%%% END FIG. 6
To better illuminate the physical structure of this enhanced geometry, 
the global potential (\ref{eq:V_global_beta}) can be alternatively written 
in the form,
\begin{align}
\label{eq:V_global_beta_alt}
V(\phi) 
& =
V_0 \left\{
\left(1 - e^{-\sqrt{\frac{2}{3}}\kappa\phi}\right)^2
\left(1- \frac{4\beta }{\frac{2}{3}\kappa^2\phi^2 + 1}\right) 
+
2\left[1 - \left( 1 + \sqrt{\frac{2}{3}}\kappa\phi\right)
e^{-\sqrt{\frac{2}{3}}\kappa\phi} \right] e^{-\sqrt{\frac{2}{3}}\kappa\phi}
\right\}
\nonumber \\
& \equiv
V_{\text{Staro}}(\phi) 
\left(1- \frac{4\beta }{\frac{2}{3}\kappa^2\phi^2 + 1}\right) 
+
2V_0 \left[1 - \left( 1 + \sqrt{\frac{2}{3}}\kappa\phi\right)
e^{-\sqrt{\frac{2}{3}}\kappa\phi} \right] e^{-\sqrt{\frac{2}{3}}\kappa\phi},
\end{align}
where 
\begin{equation}
\label{eq:VexactStarobinsky}
V_{\text{Staro}}(\phi) 
= 
V_0 \left[ 1 - \exp\left(-\sqrt{\frac{2}{3}}\kappa\phi\right) \right]^2 
\end{equation}
is the standard Starobinsky potential.
This rearrangement reveals an interesting physical architecture that 
directly mirrors the algebraic construction of the modified Jordan-frame 
Lagrangian in Eq.\ (\ref{eq:explicit_log_model}). Because the scalar 
potential is generated via the Legendre transformation, $V \propto RF - f$, 
the structurally separated negative $(y-1)^2$ sector of the Lagrangian 
maps directly to the positive primary driving term of the Einstein-frame 
potential, perfectly carrying over its rational-logarithmic modulation. 
The first term demonstrates that for large curvatures, the potential 
$V(\phi)$ is characterized by the standard Starobinsky potential 
undergoing a multiplicative suppression by a 
rational geometric regulator, becoming
\begin{equation}
\label{eq:V(phi)-UV-asymptotic}
V(\phi) 
\simeq 
V_0 \left( 1 - \frac{6\beta}{\kappa^2\phi^2} \right)
\end{equation}
in the deep inflationary ultraviolet ($\kappa\phi \gg 1$) limit
(Fig.\ \ref{fig:07}).
In this limit, the second term governed by a global $e^{-\sqrt{2/3}\kappa\phi}$ 
envelope is exponentially strongly suppressed and disappears  
from the inflationary dynamics. Conversely, in the deep infrared 
($\kappa\phi \to 0$), both terms dynamically interlock, collectively 
recovering the $V \propto \phi^2$ scalaron minimum required for a 
stable vacuum and standard post-inflationary reheating 
(Fig.\ \ref{fig:08}). 

%%%%%%%%% BEGIN FIG. 7
%%%%%%%%% END FIG. 7

%%%%%%%%% BEGIN FIG. 8
%%%%%%%%% END FIG. 8

\section{Pad\'{e}-regularized generalization}

Finally, we \emph{hypothesize} that the ``correct'' 
effective scalaron potential governing 
cosmic inflation is structurally captured by the rational Pad\'{e} [0/1] 
approximant form of the enhanced HSQRT potential $V(\phi)$ given in Eq.\ (\ref{eq:V_global_beta_alt}),
\begin{align}
\label{eq:V_global_beta_true}
V_{\text{Pad\'{e}}}(\phi) 
&=
V_0 \left\{
\frac{\left(1 - e^{-\sqrt{\frac{2}{3}}\kappa\phi}\right)^2}
{1+ \frac{4\beta }{\frac{2}{3}\kappa^2\phi^2 + 1}} 
+
2\left[1 - \left( 1 + \sqrt{\frac{2}{3}}\kappa\phi\right)
e^{-\sqrt{\frac{2}{3}}\kappa\phi} \right] e^{-\sqrt{\frac{2}{3}}\kappa\phi}
\right\}
\nonumber \\
& \equiv
\frac{V_{\text{Staro}}(\phi)} 
{1+ \frac{4\beta }{\frac{2}{3}\kappa^2\phi^2 + 1}}
+
2V_0 \left[1 - \left( 1 + \sqrt{\frac{2}{3}}\kappa\phi\right)
e^{-\sqrt{\frac{2}{3}}\kappa\phi} \right] e^{-\sqrt{\frac{2}{3}}\kappa\phi},
\end{align}
which is motivated by the standard utility of Pad\'{e} approximants in effective field 
theory, where they provide a mathematically robust method to resum perturbative 
corrections into globally regular functions that safely interpolate between 
disparate energy scales.

To fully close this logical progression (which used phenomenologically motivated 
Born-Infeld/Dirac-Born-Infeld electrodynamics, e.\ g., \cite{Sorokin2022}, 
as a conceptual starting point \cite{HSQRT}) and connect this Einstein-frame Pad\'{e} potential 
back to the corresponding (reconstructed) $f_{\text{Pad\'{e}}}(R)$ Jordan frame Lagrangian, 
we again employ an exact parametric formulation. Using the parametric variable,
\begin{equation}
y \equiv f_{\text{Pad\'{e}}}'(R) = e^{\sqrt{2/3}\kappa\phi},
\end{equation}
and the master equations,\footnote{These master identities are a direct mathematical 
consequence of the Legendre transformation mapping the Jordan frame to 
the Einstein frame. Starting from the exact definition of the scalar potential, 
$V = (R F - f)/(2\kappa^2 F^2)$, and identifying the conformal factor as 
the independent variable $y \equiv F = f'(R)$, we can rearrange the expression 
to isolate the transformed terms, $2\kappa^2 y^2 V(y) = R(y)y - f(R(y))$. 
Taking the total derivative of both sides with respect to $y$ yields 
$2\kappa^2 \frac{d}{dy}[y^2 V(y)] = \frac{dR}{dy}y + R(y) - f'(R)\frac{dR}{dy}$. 
Because the background mapping rigorously enforces $f'(R) = y$, the terms 
proportional to the implicit derivative $dR/dy$ identically cancel. This leaves 
exactly $R(y) = 2\kappa^2 \frac{d}{dy}[y^2 V(y)]$, demonstrating that the 
Jordan-frame curvature can be extracted directly from any analytic 
Einstein-frame potential without requiring transcendental inversion.}
\BEq
\label{eq:parametricRicciScalar}
R(y) = 2\kappa^2 \frac{d}{dy}[y^2 V_{\text{Pad\'{e}}}(y)],
\EEq 
and 
\BEq
\label{eq:master-eq_for-action}
f_{\text{Pad\'{e}}}(y) = y R(y) - 2\kappa^2 y^2 V_{\text{Pad\'{e}}}(y),
\EEq
we derive the Ricci scalar and the Jordan-frame Lagrangian exactly,
without analytically inverting transcendental functions,
as follows.
Substituting $\sqrt{2/3}\kappa\phi = \ln y$ into 
Eq.\ (\ref{eq:V_global_beta_true}) 
and multiplying by $y^2$, we first obtain an exact 
algebraic expression where the inverse factors of $y$ elegantly 
cancel,
\begin{equation}
y^2 V_{\text{Pad\'{e}}}(y) 
= 
V_0 \left\{ \frac{(y - 1)^2}{1 + \frac{4\beta}{\ln^2 y + 1}} + 2y - 2 - 2\ln y \right\}.
\end{equation}
Taking the derivative with respect to $y$, the parametric Ricci scalar 
(\ref{eq:parametricRicciScalar}) is then given by
\begin{equation}
\label{eq:RPade_parametric}
R(y) = 2\kappa^2 V_0 \left\{ 
\frac{2(y-1)}{1 + \frac{4\beta}{\ln^2 y + 1}} 
+ \frac{8\beta(y-1)^2 \ln y}{y \left[ \ln^2 y + 1 + 4\beta \right]^2} 
+ \frac{2(y-1)}{y} \right\}.
\end{equation}
Applying the master equation for the action (\ref{eq:master-eq_for-action}),
the resulting terms simplify, with the numerators combining 
as $2y^2 - 2y - y^2 + 2y - 1 = y^2 - 1$, yielding the exact, 
closed-form parametric Lagrangian for our modified gravity model,
\begin{equation}
\label{eq:fyPade_parametric}
f_{\text{Pad\'{e}}}(y) = 2\kappa^2 V_0 \left\{ 
\frac{y^2 - 1}{1 + \frac{4\beta}{\ln^2 y + 1}} 
+ \frac{8\beta(y-1)^2 \ln y}{\left[ \ln^2 y + 1 + 4\beta \right]^2} 
+ 2\ln y \right\}.
\end{equation}

This exact parametric pair, $\{R(y), f_{\text{Pad\'{e}}}(y)\}$, constitutes 
a complete specification of the higher-derivative Lagrangian. 
We can immediately perform two important structural 
checks. First, we examine the Pad\'{e} denominator for phantom poles. 
Because the ultraviolet coupling $\beta > 0$ and the squared logarithm 
$\ln^2 y \geq 0$, the denominator $1 + 4\beta / [\ln^2 y + 1]$ is strictly 
greater than or equal to unity. The modified potential is therefore unconditionally 
stable and globally smooth, entirely circumventing the phantom pole singularities 
that typically plague rational reconstructions.

Second, we must ensure that the rational ultraviolet modification does not 
inadvertently destroy the infrared ($y \rightarrow 0$) architecture originally 
designed to cure the $f'(R)=0$ singularity of the standard Starobinsky model. 
In the deep infrared limit ($y \rightarrow 0$), the logarithm diverges 
($\ln y \rightarrow -\infty$), causing the rational modification in the 
denominator of Eq.\ (\ref{eq:V_global_beta_true}) to vanish, 
$\frac{4\beta}{\ln^2 y + 1} \rightarrow 0$. The entire rational term reduces to 
a harmless constant, such that the function simplifies asymptotically to 
$y^2 V_{\text{Pad\'{e}}}(y) \approx V_0(-1 - 2\ln y)$. 
Evaluating the curvature via the derivative yields 
$R(y) \approx -4\kappa^2 V_0 / y$. 
Consequently, as $y \rightarrow 0$, the Ricci scalar correctly diverges to 
$R \rightarrow -\infty$. The vital logarithmic term responsible for pushing the 
$f'(R)=0$ crossing infinitely far away in the base HSQRT model is left 
completely undisturbed, proving that the Pad\'{e}-regularized enhanced model 
successfully engineers the desired large-field ultraviolet plateau while 
rigorously preserving its singularity-free infrared completion.

\section{Cosmological Observables and 2025--2026 Observational Constraints}

With the explicit $1/\phi^2$ asymptotic potential, 
Eq.\ (\ref{eq:V(phi)-UV-asymptotic}), established in the Einstein 
frame, the quantum fluctuations of the scalaron 
are governed by the canonical quantization framework 
\cite{Sasaki1986, LythRiotto1999}. To rigorously justify the 
use of the asymptotic limit, we must first mathematically 
estimate the canonical field values during the inflationary 
epoch.\footnote{To clarify the standard cosmological timeline 
and terminology used to describe the scalar field trajectory: 
The duration of the inflationary expansion is measured by 
the number of e-folds, $N = \ln(a_{\text{final}} / a_{\text{initial}})$, 
conventionally counted backward from the end of the epoch 
($N=0$). The ``end of inflation'' is defined as the exact moment 
the potential steepens sufficiently for the slow-roll approximation 
to break down ($\epsilon_V \simeq 1$). In our enhanced geometry, 
this corresponds to $\kappa\phi_{\text{end}} \approx 0.84$, 
as derived in Eq.\ (\ref{eq:xend}), after which the scalaron falls into 
the potential minimum to initiate reheating. 
``Horizon exit'' refers to the physical process 
whereby the wavelengths of microscopic quantum fluctuations 
are stretched exponentially until they exceed the causal Hubble 
horizon, freezing out as macroscopic, classical density perturbations. 
The ``observable window'' corresponds to the specific period 
when the perturbations responsible for the large-scale structure 
of the present-day cosmic microwave background exited the 
horizon, generally occurring $N \simeq 50{-}60$ e-folds prior 
to the end of inflation. In the present model, $N \simeq 55$ 
corresponds to the field traversing the region 
$\kappa\phi_* \sim 5.0$, approximately where the inverse-power 
rational tail overtakes the standard Starobinsky plateau 
(Fig.\ \ref{fig:07}).} 

The end of inflation is demarcated by the breakdown of the slow-roll 
approximation, $\epsilon_V \simeq 1$. Because the rational $\beta$-regulator 
vanishes in the deep infrared, the potential reverts 
to the base HSQRT geometry,
\begin{equation}
V(\phi) 
\simeq 
V_{\text{base}}(\phi)
=
V_0 \left[ 1 - \left(1+2\sqrt{\frac{2}{3}}\kappa\phi\right)
e^{-2\sqrt{\frac{2}{3}}\kappa\phi} \right].
\end{equation}
Evaluating the first slow-roll parameter, 
$\epsilon_V \equiv (1/2\kappa^2)(V_{,\phi}/V)^2 \simeq 1$, yields a 
transcendental equation for the dimensionless field 
$x_{\text{end}} \equiv \sqrt{2/3}\kappa\phi_{\text{end}}$,
\begin{equation}
\label{eq:xend}
\frac{4 x_{\text{end}} e^{-2x_{\text{end}}}}
{\sqrt{3} \left[ 1 - (1+2x_{\text{end}})e^{-2x_{\text{end}}} \right]} 
= 1.
\end{equation}
Solving this numerically gives $x_{\text{end}} \approx 0.69$, corresponding to an 
end-of-inflation field value of $\kappa\phi_{\text{end}} \approx 0.84$. 

The observable cosmic microwave background (CMB) fluctuations, however, 
exit the horizon $N \simeq 55$ e-folds prior to the end of inflation. 
Because the total potential gradient in the intermediate region 
is a superposition of the decaying exponential slope and the emerging 
rational tail, integrating the exact number of e-folds requires 
evaluating the full potential, $N \simeq \int (V/V_{,\phi}) d\phi$. 
Since the gradients combine additively 
($V_{,\phi}^{\text{total}} = V_{,\phi}^{\text{base}} + V_{,\phi}^{\text{corr}}$), 
the field traverses the intermediate geometry faster than in the unperturbed 
model. Numerical evaluation for $\beta \sim 0.1$ and $N=55$  
places horizon exit at $\kappa\phi_* \approx 5.0$. 

At this scale of horizon exit, the global exponential envelope is heavily 
suppressed, reducing the base exponential gradient to a sub-leading correction 
of the total force driving the scalaron. Consequently, the rational inverse-power tail 
dominates the slow-roll hierarchy during the observable window. 
Further up the plateau ($\kappa\phi \gtrsim 5.75$), the exponentially decaying 
transient terms become asymptotically negligible. Because inverse polynomials 
decay much slower than exponentials, the $1/\phi^2$ term fully determines
the asymptotic behavior of the potential at large field values.

We now derive the primordial observables generated by this enhanced geometry. 
In this deep ultraviolet regime, the leading-order potential derivatives are
\BEq
V_{,\phi} \simeq \frac{12 \beta V_0}{\kappa^2 \phi^3},
\quad
V_{,\phi\phi} \simeq -\frac{36\beta V_0}{\kappa^2 \phi^4}.
\EEq 
Consequently, the standard potential slow-roll parameters, defined as 
\BEq
\epsilon_V = \frac{1}{2\kappa^2}\left(\frac{V_{,\phi}}{V}\right)^2, 
\quad 
\eta_V = \frac{1}{\kappa^2}\frac{V_{,\phi\phi}}{V},
\EEq
evaluate directly to
\begin{equation}
\epsilon_V(\phi) \simeq \frac{72 \beta^2}{\kappa^6 \phi^6} , \quad
\eta_V(\phi) \simeq -\frac{36 \beta}{\kappa^4 \phi^4}.
\end{equation}
The number of e-folds $N$ from the end of inflation is obtained 
by integrating the inverse of the first derivative,
\begin{equation}
N \simeq \kappa^2 \int \frac{V}{V_{,\phi}} \, d\phi 
\simeq \kappa^2 \int \frac{\kappa^2 \phi^3}{12 \beta} \, d\phi 
\simeq \frac{\kappa^4 \phi^4}{48 \beta}.
\end{equation}
Inverting this relation yields the mapping 
$\kappa^4 \phi^4 \simeq 48 \beta N$, which implies 
$\kappa^6 \phi^6 \simeq 192\sqrt{3} \beta^{3/2} N^{3/2}$. 
Substituting this scaling back into the slow-roll parameters produces 
a hierarchical scaling where the parameter $\beta$ exactly cancels 
out of $\eta_V$,
\begin{equation}
\epsilon_V(N) \simeq \frac{72 \beta^2}{192\sqrt{3} \beta^{3/2} N^{3/2}} 
= \frac{(3 \beta)^{1/2}}{8 N^{3/2}}, \quad
\eta_V(N) \simeq -\frac{36 \beta}{48 \beta N} = -\frac{3}{4 N}.
\end{equation}

Because $\epsilon_V \propto N^{-3/2}$ and $\eta_V \propto N^{-1}$, 
the first slow-roll parameter is parametrically suppressed relative to 
the second for large e-foldings ($N \sim 50{-}60$) and small UV couplings 
($\beta \ll 1$). Consequently, the scalar spectral index is overwhelmingly 
dominated by $\eta_V$, yielding a universal, parameter-free prediction 
at leading order,
\begin{equation}
\label{eq:ns_prediction}
n_s \simeq 1 - 6\epsilon_V + 2\eta_V \simeq 1 - \frac{3}{2 N}.
\end{equation}
For the standard window of inflationary duration, $N \in [50, 60]$, 
Eq.~(\ref{eq:ns_prediction}) bounds the scalar spectral index to 
the domain $n_s \in [0.970, 0.975]$. This theoretical prediction is 
of paramount importance. While the unperturbed conformal trap 
yields $n_s \simeq 0.960{-}0.966$, the integration of high-resolution 
ground-based CMB data (ACT DR6) and recent baryon acoustic 
oscillation measurements (DESI) has systematically shifted 
the favored central value of the spectral tilt upward, tightening 
the bounds around $n_s \gtrsim 0.97$. The logarithmic enhancement 
naturally drives the spectral index directly into this newly favored 
observational window.\footnote{To contextualize these predictions 
within the broader landscape of inflationary models, it is instructive 
to map our Einstein-frame potentials onto the systematic classification 
provided by the \textit{Encyclop\ae dia Inflationaris} \cite{Martin2014}. 
The baseline HSQRT geometry yields a potential that approaches 
the inflationary plateau via an exponential suppression, 
$V(\phi) \simeq V_0 (1 - \mathcal{O}(\phi) e^{-\gamma \phi})$. 
This behavior places the unperturbed baseline model within 
the universality class of standard plateau paradigms, together 
with models such as Higgs-Starobinsky Inflation (HI) and K\"ahler 
Moduli Inflation I (KMII). These models are characterized by a universal 
leading-order spectral index of $n_s \simeq 1 - 2/N$. Conversely, 
the logarithmically enhanced HSQRT model explicitly breaks 
this exponential attractor. As derived above, the resulting potential 
approaches the plateau via an inverse-square law, 
$V(\phi) \simeq V_0 (1 - 6 \beta/(\kappa^2 \phi^2))$. Within the classification 
of Ref.\ \cite{Martin2014}, this puts it into the Brane Inflation (BI) 
class, governed by the generic asymptotic form 
$V(\phi) \propto 1 - (\mu/\phi)^p$. For this specific class of potentials, 
the leading-order spectral index is modified to 
$n_s \simeq 1 - \frac{2(p+1)}{p+2} \frac{1}{N}$, which for $p=2$ 
yields $n_s \simeq 1 - 3/(2N)$. Therefore, the functional shift from 
an exponential to an inverse-power approach effectively pushes  
the geometry from the HI/KMII universality class into the BI ($p=2$) 
class. This structural transition provides the precise mathematical 
mechanism by which the enhanced geometry elevates the spectral 
index to accommodate current observational constraints.}

Next, we evaluate the running of the spectral index, 
$\alpha_s = d n_s / d \ln k$. In standard single-field slow-roll scenarios, 
the running is inherently a second-order effect. Differentiating the 
leading-order spectral index $n_s \simeq 1 - 3/(2N)$ with respect to 
the e-folding number, $d\ln k \simeq -dN$, yields the analytic prediction,
\begin{equation}
\label{eq:alphas_prediction}
\alpha_s \simeq -\frac{d n_s}{d N} \simeq -\frac{3}{2 N^2}.
\end{equation}
Evaluating this across the standard $N \in [50, 60]$ window produces 
an exceptionally small, negative running, $\alpha_s \in [-0.00060, -0.00042]$. 
We compare this against the recent high-resolution constraints from 
the ACT DR6, which reports $\alpha_s = 0.0062 \pm 0.0052$ \cite{ACT:2025-2}. 
While the central value of the ACT data exhibits a slight preference 
for positive running, our theoretical prediction sits comfortably within the $2\sigma$ 
confidence interval ($\sim 1.3\sigma$ deviation). The exceedingly small, 
negative magnitude of $\alpha_s \sim \mathcal{O}(10^{-4})$ is a universal 
hallmark of canonical single-field models governed by large-$N$ asymptotics. 
A statistically significant detection of $\mathcal{O}(10^{-3})$ positive 
running would inherently require a departure from the standard slow-roll 
hierarchy. Thus, the enhanced HSQRT model remains thoroughly consistent 
with the current bounds on $\alpha_s$, pending tighter constraints from 
forthcoming CMB surveys.

Finally, the amplitude of primordial gravitational waves is quantified by 
the tensor-to-scalar ratio, $r \simeq 16\epsilon_V$. Substituting the explicit 
scaling for $\epsilon_V(N)$ yields,
\begin{equation}
\label{eq:r_prediction}
r \simeq \frac{2(3 \beta)^{1/2}}{N^{3/2}}.
\end{equation}
This explicit $\beta$-dependence fundamentally breaks the restrictive 
consistency relation $r \propto (1-n_s)^2$ inherent to purely exponential 
potentials. The transition to the $1/\phi^2$ inverse-power potential softens 
the asymptotic suppression of the tensor signal from the stringent 
Starobinsky scaling ($1/N^2$) to a gentler scaling ($1/N^{3/2}$). 
More importantly, while $n_s$ is anchored to the observational sweet 
spot by the universal $1/N$ relation, the tensor-to-scalar ratio $r$ remains 
a fully tunable target. By selecting $\beta \ll 1$, the model easily satisfies 
the stringent upper bounds from the latest joint analyses ($r < 0.03$), 
while retaining the phenomenological flexibility to produce an observable 
B-mode polarization signal for next-generation observatories.

\section{Scalaron Dynamics and Post-Inflationary Reheating}

Following the termination of the slow-roll phase, the universe must 
gracefully exit the inflationary de Sitter attractor and transition into a 
standard radiation-dominated epoch. In $f(R)$ gravity theories, this 
thermalization is driven by the coherent oscillations of the massive 
scalaron field around the global minimum of its effective Einstein-frame 
potential, followed by its perturbative decay into Standard Model 
degrees of freedom. To verify the viability of this phase, it is imperative 
to examine the deep infrared limit ($R \to 0$) of the modified geometry 
to ensure that standard general relativity is cleanly recovered and that 
the rational-logarithmic enhancement does not inadvertently destabilize 
the classical vacuum.

\subsection{Vacuum Stability and the Effective Equation of State}

In the unperturbed baseline model ($\beta = 0$), the cessation of 
inflation is characterized by the scalar curvature dropping toward 
zero, corresponding to the parametric limit $y \to 1$. Expanding 
the baseline Lagrangian around this vacuum yields the linear 
general relativity limit equipped with a quadratic correction,
\begin{equation}
f_{\text{base}}(R) \simeq R + \frac{\alpha}{2} R^2 + \mathcal{O}(R^3).
\end{equation}
Consequently, the baseline geometry successfully establishes a stable 
vacuum where $f'(0) = 1$ and $f''(0) = \alpha$. In the Einstein frame, 
expanding the exact potential $V(y)$ and the canonical field $\phi(y)$ 
around $y=1$ generates a perfectly harmonic minimum,
\begin{equation}
V(\phi) \simeq \frac{1}{2} m_s^2 \phi^2, 
\quad 
m_s^2 = \frac{1}{3 f''(0)} = \frac{1}{3\alpha}.
\end{equation}
Because the potential maintains this quadratic shape at the 
bottom of the well, the coherent oscillations of the scalaron field 
dynamically average to an effective equation of state $w \simeq 0$. 
During this entire post-inflationary epoch, the scalaron fluid acts 
as a cold, pressureless dust component before eventually decaying.

\subsection{Amplified Thermalization in the Enhanced Geometry}
The true physical robustness of the model becomes apparent when 
analyzing the exact enhanced geometry. It is a notoriously common 
pathology that non-perturbative or rational corrections designed to 
modify the deep ultraviolet plateau inadvertently destroy the $R \to 0$ 
limit, introducing phantom poles or shifting the vacuum state. However, 
in the enhanced HSQRT geometry (\ref{eq:explicit_log_model_R}), 
the logarithmic modification asymptotically vanishes in the infrared, producing 
\BEq
m_s^2(R)
=
\frac{1}{3 \alpha  (1+2 \beta)}
-\frac{(1+6 \beta) R}{3 (1+2 \beta)^2}
+\frac{\left(1 +32 \beta + 80 \beta ^2 - 24 \beta ^3 \right) \alpha R^2 }
{6 (1+2 \beta)^3}+{\cal O}\left(R^3\right),
\EEq
and leaving the global minimum strictly intact, with
the zero-curvature mass being
\begin{equation}
m_{s}^2(0) = \frac{1}{3\alpha(1 + 2\beta)}.
\end{equation}

Comparing this against the standard Starobinsky case
($m_{\text{Staro}}^2 = 1/(6\alpha)$), we see that the baseline HSQRT 
mass is naturally amplified by exactly a factor of two. Because our 
deformation parameter is observationally required to be perturbative 
($\beta \sim \mathcal{O}(10^{-1})$) to ensure the rational tail governs 
the observable CMB window,\footnote{The specific parameter choice 
$\beta \sim \mathcal{O}(10^{-1})$ represents a phenomenological sweet spot 
for this enhanced geometry. If the coupling were excessively suppressed 
(e.g., $\beta \sim 10^{-2}$), horizon exit at $N \simeq 55$ would occur too far 
down the plateau ($\kappa\phi_* \approx 2.3$), where the baseline exponential terms 
are no longer negligible. In this mixed regime, the clean analytic predictions 
derived from the pure $1/\phi^2$ asymptotic tail would break down. Conversely, 
$\beta \sim 0.1$ pushes horizon exit to $\kappa\phi_* \approx 5.0$, ensuring the 
exponentially decaying transient terms are heavily suppressed 
($e^{-2\sqrt{2/3}\kappa\phi_*} \sim \mathcal{O}(10^{-4})$), validating 
the inverse power approximation. Furthermore, this value predicts a tensor-to-scalar 
ratio of $r \approx 0.003$, safely an order of magnitude below current 
observational upper bounds.} the enhanced mass $m_{s}(0)$ only experiences a 
fractional $\mathcal{O}(\beta)$ shift from this amplified baseline. This ensures that 
the local Compton wavelength of the scalaron remains tightly bound, 
safely satisfying all macroscopic fifth-force and solar system constraints.

Finally, this geometric mass amplification yields a distinct phenomenological 
signature for the thermal history of the universe. The scalaron decays into 
conformally coupled Standard Model fields with a perturbative decay rate 
scaling as $\Gamma \propto \kappa^2 m_s^3$. Consequently, the amplified 
baseline mass inherently yields a decay rate larger than the standard 
Starobinsky case by a factor of $2\sqrt{2}$. Because the reheating 
temperature scales as $T_{\text{re}} \sim 0.1 \sqrt{\Gamma / \kappa}$, 
this shifts the baseline temperature upward by a factor of $2^{3/4} \simeq 1.68$, 
while the logarithmic enhancement subsequently modulates it by a modest 
factor of $(1+2\beta)^{-3/4}$. 

For $\beta \ll 1$, this modulation is cosmologically safe. The explicitly 
formulated logarithmically enhanced deformation successfully governs the 
extreme ultraviolet approach to the inflationary plateau without corrupting 
the infrared vacuum stability, leaving the reheating temperature firmly 
anchored within the phenomenologically optimal 
$10^8 \text{--} 10^9 \text{ GeV}$ window. This securely precedes Big Bang 
Nucleosynthesis (BBN) while remaining low enough to strictly avoid the 
late-time overproduction of thermal gravitinos.

\section{The Negative Curvature Regime and the Strong-Coupling Boundary}
A defining characteristic of the baseline hyperbolic square-root geometry 
is its behavior in the limit of extreme negative curvature, $R \to -\infty$. 
In standard quadratic gravity, this limit forces $f'(R) < 0$, plunging the 
theory into a tachyonic ghost regime. In contrast, the baseline HSQRT 
model restricts the conformal factor to strictly positive values, 
$F_{\text{base}} \to 0^+$, erecting an infinite energetic wall in the Einstein 
frame that dynamically forbids access to the pathological domain. 
For the enhanced model to remain globally viable, the introduced logarithmic 
enhancement must not disrupt this protective asymptotic structure.

In our exact parametric formulation, the strong-coupling boundary 
$R \to -\infty$ corresponds to the limit $y \to 0^+$. We must rigorously 
evaluate the behavior of the enhanced conformal factor $F(y)$ in this 
regime to determine if the proposed enhancement artificially forces 
the geometry into the ghost domain.

Recall the exact form of the conformal factor,
\[
F(y) = y + \frac{2\beta y^2}{y^2 + 1} H'(y),
\]
where the derivative of the correction function is
\[
H'(y) = \frac{2(y - 1)}{\ln^2 y + 1} \left[ 1 - \frac{(y - 1)\ln y}{y(\ln^2 y + 1)} \right].
\]
As $y \to 0^+$, the transcendental terms dominate. The logarithmic 
functions diverge as $\ln y \to -\infty$ and $\ln^2 y \to +\infty$. 
Expanding the dominant terms inside the bracket of $H'(y)$, 
the expression asymptotically scales as
\begin{equation}
H'(y) \simeq \frac{-2}{\ln^2 y} \left[ \frac{-\ln y}{y \ln^2 y} \right] 
\simeq \frac{-2}{y \ln^3 y}.
\end{equation}
Substituting this asymptotic scaling back into the correction term 
for the conformal factor, we find,
\begin{equation}
\Delta F(y) \equiv \frac{2\beta y^2}{y^2 + 1} H'(y) 
\simeq 2\beta y^2 \left( \frac{-2}{y \ln^3 y} \right) 
= -\frac{4\beta y}{\ln^3 y}.
\end{equation}
Because $y$ approaches zero much faster than any inverse power 
of $\ln y$, the physical limit evaluates to zero,
\begin{equation}
\lim_{y \to 0^+} \Delta F(y) = 0.
\end{equation}
Consequently, in the deep negative curvature regime, 
the enhanced conformal factor reduces to the baseline limit,
\begin{equation}
\lim_{y \to 0^+} F(y) = y \to 0^+.
\end{equation}
Because $F(y)$ remains strictly positive, the condition $f'(R) > 0$ 
is globally preserved, ensuring the extra scalar degree of freedom 
never becomes a ghost. 

Let us also evaluate the Einstein-frame potential $V(y)$ at this 
boundary. As $y \to 0^+$, the canonical field diverges to negative 
infinity, $\phi \propto \ln y \to -\infty$. The potential is governed 
by $V(y) = [R(y)F(y) - f(y)] / [2\kappa^2 F(y)^2]$. While the $R F$ 
term approaches a finite constant $-1/(2\alpha)$, the baseline 
Lagrangian diverges due to the logarithmic component 
$f(y) \sim \frac{\ln y}{2\alpha} \to -\infty$. Since 
$F(y)^2 \to y^2 \to 0^+$, the total potential scales as,
\begin{equation}
V(y) \simeq \frac{-\ln y / (2\alpha)}{2\kappa^2 y^2} \to +\infty.
\end{equation}

This confirms that the explicit logarithmic deformation asymptotically 
vanishes at the strong-coupling boundary. The infinite energetic 
wall in the Einstein frame is perfectly preserved, dynamically freezing 
out the scalaron and forbidding access to the tachyonic regime at 
extreme negative curvatures.

\section{Discussion}

With the mathematical machinery and observational predictions of the enhanced model 
now fully established, it is instructive to step back and examine the structural 
necessity of this specific construction. The transition from a strict exponential 
to an inverse-power inflationary plateau intrinsically requires modifications to 
the underlying $f(R)$ action. While a rigorous, top-down derivation of the exact 
global effective action from quantum gravity remains beyond current theoretical 
reach, the proposed framework can be understood as a mathematically robust, 
bottom-up phenomenological ansatz. The objective of this formulation is not to 
claim a fundamental Lagrangian, but to provide an explicit proof-of-concept 
demonstrating how observationally favored inverse-power plateaus can be globally 
stabilized in $f(R)$ gravity.

\subsection{EFT Motivation and the Bottom-Up Ansatz}

In the Einstein frame, forcing the scalar potential to approach its plateau via an 
inverse-power law, $V(\phi) \simeq V_0 (1 - \mu / \phi^2)$, imposes a specific mathematical 
requirement on the geometric derivative: $f'(R)$ must exhibit a logarithmic dependence 
at large curvatures. 

This functional behavior is broadly motivated by effective field theory (EFT) and 
semiclassical gravity \cite{QuantumGravity1, QuantumGravity2}. When conformal matter 
fields are integrated out on a highly curved background, the resulting 1-loop 
effective action inevitably acquires higher-order curvature terms. Manifesting 
prominently through the conformal trace anomaly, these quantum contributions typically 
appear as operators proportional to $R^2 \ln (R/\mu^2)$ \cite{Barvinsky2023a, 
Barvinsky2023, Cognola2005, Mirzabekian1996}. Similarly, functional 
renormalization-group (FRG) studies suggest that the effective action naturally 
acquires terms that run logarithmically with the background curvature scale 
\cite{Bonanno2002RGInflation, Reuter2012AsymptoticSafety, Alexandre2025RGFlowACT}. 

Because the exact resummation of these terms into a complete global action is 
unknown, cosmological model building often relies on targeted deformations. 
In our formulation, the rational-logarithmic enhancement of the HSQRT deformation 
is parameterized by the dimensionless coupling $\beta$, which serves as 
an effective macroscopic parameter governing the scale of the inverse-power tail.

\subsection{The Branch-Cut Pathology of Naive Logarithms}

The primary obstacle in phenomenological $f(R)$ model building is that directly appending 
a raw logarithm to a standard quadratic action, such as writing 
$f(R) = R + \alpha R^2 + \beta R^2 / \ln^2(\alpha R)$, introduces  
complex singularities. The structural vulnerability lies in the fact that the scalar 
curvature $R$ is not strictly positive-definite throughout cosmic history. 
While $R > 0$ during inflation, the Ricci scalar vanishes ($R \simeq 0$) during 
radiation domination, which causes a raw logarithmic denominator to diverge. 
Furthermore, $R$ may drop below zero during pre-inflationary dynamics, bouncing 
cosmologies, or near the strong-coupling boundary ($R \to -\infty$). 

The moment the scalar curvature becomes negative, a term like $\ln(\alpha R)$ generates 
a branch cut. The physical action becomes complex, the canonical scalar field acquires 
imaginary components, and the potential $V(\phi)$ becomes ill-defined. This analytic 
obstruction severely limits the ability to safely embed logarithmically running terms 
into standard $f(R)$ theories.

\subsection{The HSQRT Geometry as a Mathematical Regularizer}

To bypass this branch-cut pathology, an algebraic regularization is required. Rather 
than treating the baseline HSQRT geometry as a fundamental truth, we employ it here 
as a mathematically necessary scaffold. It serves as an algebraic regularizer that 
allows the required logarithmic dependencies to exist globally without compromising 
the real-valued nature of the action.

The defining feature of the baseline HSQRT model is the replacement of the linear 
derivative term with the exact hyperbolic mapping,
\begin{equation}
y(R) = \alpha R + \sqrt{\alpha^2 R^2 + 1}.
\end{equation}
This specific mapping serves an important stabilizing function. It compresses 
the entire infinite real line of scalar curvature, $R \in (-\infty, +\infty)$, strictly 
into the positive real domain, $y \in (0, +\infty)$. 

By injecting the logarithmic terms explicitly as functions of the parametric 
variable $y$ rather than the raw curvature $R$ (utilizing $\ln y$ instead of 
$\ln R$), the argument of the logarithm is strictly positive-definite across all curvature 
scales. Even in the extreme negative curvature limit ($R \to -\infty$), the variable $y$ 
remains positive, ensuring the logarithmic terms never cross a branch cut. 

In our approach, the enhanced HSQRT geometry operates as an explicit existence proof. 
It demonstrates that it is mathematically possible to embed the scale-dependent logarithmic 
running required for an inverse-power plateau into a globally real-valued action. By 
rigorously protecting the background from tachyonic and ghost pathologies at all scales, 
this ansatz provides a stable phenomenological host for the gravitational instability of 
the scalaron field \cite{Mukhanov1985}, allowing for a consistent derivation of 
the modified inflationary observables.

Ultimately, mathematical consistency is only a prerequisite for an acceptable physical theory. 
While the enhanced HSQRT model demonstrates that a globally stable, inverse-power 
inflationary plateau can be explicitly constructed, algebraic viability does not 
guarantee physical reality. Whether such scale-dependent logarithmic running and 
its corresponding inflationary predictions are actually realized in Nature is a 
question that cannot be answered by formal proofs of concept, but only by actual 
observations. Future high-precision measurements of the cosmic microwave background, 
particularly tighter constraints on the tensor-to-scalar ratio, will ultimately 
determine whether this phenomenological mechanism correctly describes the early 
universe.

\section{Conclusion}

In this paper, we have introduced and analyzed a logarithmically 
enhanced extension of the hyperbolic square-root (HSQRT) $f(R)$ 
inflationary model. The primary motivation for this development is 
the emerging preference in recent high-precision joint CMB and 
Large-Scale Structure datasets for a scalar spectral index 
slightly higher than the predictions of pure exponential plateau models. 
By incorporating a functionally regularized logarithmic deformation 
governed by a single ultraviolet parameter $\beta$, we established 
a theoretical mechanism that smoothly transitions the asymptotic 
behavior of the Einstein-frame potential from a strict exponential 
to an inverse-power scaling.

A central methodological result of this paper is the exact parametric 
formulation of the modified geometry. Because the logarithmic 
deformation inherently breaks the algebraic invertibility of the 
conformal mapping, we expressed the complete dynamics of the 
Einstein frame via a parametric system governed by the 
derivative variable $y$. This approach provides a globally continuous, 
mathematically exact system that facilitates analytic expansions in all 
critical cosmological limits without relying on piecewise approximations 
of the scalar potential.

Phenomenologically, the enhanced model demonstrates excellent 
alignment with current observational constraints. The transition 
to an inverse-power plateau scaling dynamically shifts the geometry into 
the $p=2$ Brane Inflation (BI) universality class. This drives the leading-order 
prediction for the scalar spectral index to $n_s \simeq 1 - 3/(2N)$, strictly 
anchoring it within the observationally favored window of $n_s \in [0.970, 0.975]$ 
for standard inflationary durations. Additionally, the tensor-to-scalar ratio 
asymptotically scales as $r \simeq 2(3\beta)^{1/2}/N^{3/2}$. This explicit 
$\beta$-dependence explicitly breaks the consistency relation 
inherent to unperturbed conformal attractors, allowing $r$ to act as a 
tunable observable that comfortably satisfies current upper bounds 
while remaining within the sensitivity thresholds of next-generation 
cosmic microwave background observatories.

Furthermore, we rigorously verified the theoretical integrity of the 
model across extreme curvature scales. In the infrared limit ($R \to 0$), 
the rational-logarithmic structure is asymptotically suppressed, perfectly 
recovering Einstein's general relativity ($f'(0)=1$) and preserving a stable, 
harmonic vacuum that ensures a standard reheating history. 
In the deep negative curvature regime ($R \to -\infty$), 
the asymptotic behavior of the deformation term 
safely vanishes, preserving the globally positive conformal factor 
($f'(R) > 0$) and the divergent energetic barrier that dynamically protects 
the theory from tachyonic ghost instabilities. 

Conceptually, the explicit mathematical form of the deformation is inspired 
by the structure of Pad\'{e}-type resummations, which are routinely required 
to regulate the logarithmic divergences expected from the conformal trace 
anomaly and functional renormalization-group flows. Rather than a rigorous 
first-principles derivation, the $\beta$-modulated term serves as a mathematically 
exact phenomenological ansatz that effectively mimics these scale-dependent 
semiclassical EFT expectations at the inflationary scale, using the HSQRT 
baseline as an indispensable algebraic regularizer against branch-cut pathologies.

Such a geometrically regularized logarithmic approach is highly synergistic 
with recent developments in the literature. Motivated by the exact same 
observational tensions in the ACT DR6 and DESI DR2 data, recent works have 
explored marginal quantum-motivated deformations \cite{Yuennan2025} 
or higher-order curvature additions \cite{Addazi2025, Bezerra-Sobrinho2025} 
to rescue the Starobinsky attractor. Furthermore, structurally similar 
logarithmic deformations in $f(R)$ gravity have recently been shown to 
successfully unify early-universe inflation with realistic late-time dynamical 
dark energy, providing a superior fit to the latest DESI and Pantheon+ 
observations compared to standard $\Lambda$CDM \cite{Odintsov2026}. 
This suggests that logarithmic modulations of the spacetime geometry 
may be a universal and necessary feature for reconciling modified gravity 
with the next decade of precision cosmology across both the deep ultraviolet 
and deep infrared limits.

Several directions remain open for future investigation. While the preservation 
of the infinite potential wall ensures classical stability in the Einstein frame, 
a comprehensive analysis of the corresponding non-singular bouncing 
mechanics ($H_J = 0$, $\dot{H}_J > 0$) requires a complete treatment of 
the coupled fluid dynamics and scale factor transformations within 
the Jordan frame \cite{Kamenshchik2016}. Additionally, while our $\beta$ 
parameter currently acts as an effective macroscopic coupling, formally 
deriving its specific value from exact functional flow equations within 
the Asymptotic Safety program remains a compelling goal. Finally, investigating 
whether these globally regularized conformal mappings can be formally 
linked to the gradient dynamics of Ricci flows and Perelman's entropy 
functional \cite{Hamilton1982, Perelman2002, Friedan1980, Friedan1985}
represents a fascinating theoretical avenue to further clarify the deep geometric 
origins of the observed spectral tilt.


\begin{thebibliography}{99}
\bibitem{Guth1981}
A.\ H.\ Guth,
Inflationary universe: A possible solution to the horizon and flatness problems, 
Phys.\ Rev.\ D \textbf{23}, 347 (1981).

\bibitem{Linde1982}
A.\ D.\ Linde,
A new inflationary universe scenario: A possible solution of the horizon, 
flatness, homogeneity, isotropy and primordial monopole problems, 
Phys.\ Lett.\ B \textbf{108}, 389 (1982).

\bibitem{AlbrechtSteinhardt1982}
A.\ Albrecht and P.\ J.\ Steinhardt,
Cosmology for Grand Unified Theories with Radiatively Induced Symmetry Breaking, 
Phys.\ Rev.\ Lett.\ \textbf{48}, 1220 (1982).

\bibitem{Linde1983}
A.\ D.\ Linde,
Chaotic inflation, 
Phys.\ Lett.\ B \textbf{129}, 177 (1983).

\bibitem{Linde:1984ir} 
A.\ D.\ Linde,
The Inflationary Universe,
Rept.\ Prog.\ Phys.\  {\bf 47}, 925 (1984).

\bibitem{Hawking1982}
S.\ W.\ Hawking,
The development of irregularities in a single bubble inflationary universe,
Phys.\ Lett.\ B \textbf{115}, 295 (1982).

\bibitem{GuthPi1982}
A.\ H.\ Guth and S.-Y.\ Pi,
Fluctuations in the New Inflationary Universe, 
Phys.\ Rev.\ Lett.\ \textbf{49}, 1110 (1982).

\bibitem{BardeenSteinhardtTurner1983}
J.\ M.\ Bardeen, P.\ J.\ Steinhardt and M.\ S.\ Turner,
Spontaneous creation of almost scale-free density perturbations 
in an inflationary universe,
Phys.\ Rev.\ D \textbf{28}, 679 (1983).

\bibitem{Mukhanov1985}
V.\ F.\ Mukhanov,
Gravitational instability of the universe filled with a scalar field,
JETP Lett.\ \textbf{41}, 493 (1985).

\bibitem{Sasaki1986}
M.\ Sasaki,
Large scale quantum fluctuations in the inflationary universe,
Prog.\ Theor.\ Phys.\ \textbf{76}, 1036 (1986).

\bibitem{LythRiotto1999}
D.\ H.\ Lyth and A.\ Riotto,
Particle physics models of inflation and the cosmological density 
perturbation, 
Phys.\ Rept.\ \textbf{314}, 1 (1999);
arXiv:hep-ph/9807278.

\bibitem{BassettTsujikawaWands2006}
B.\ A.\ Bassett, S.\ Tsujikawa and D.\ Wands,
Inflation dynamics and reheating,
Rev.\ Mod.\ Phys.\ \textbf{78}, 537 (2006);
arXiv:astro-ph/0507632.

\bibitem{Baumann2009}
D.\ Baumann,
TASI lectures on inflation, 
arXiv:0907.5424 [hep-th].

\bibitem{GuthKaiserNomura2014}
A.\ H.\ Guth, D.\ I.\ Kaiser and Y.\ Nomura,
Inflationary paradigm after Planck 2013, 
Phys.\ Lett.\ B \textbf{733}, 112 (2014).

\bibitem{MukhanovBook2005}
V.\ Mukhanov,
\textit{Physical Foundations of Cosmology}
(Cambridge University Press, 2005).

\bibitem{Weinberg2008}
S.\ Weinberg,
\textit{Cosmology}
(Oxford University Press, 2008).

\bibitem{Starobinsky1980}
A.\ A.\ Starobinsky,
A new type of isotropic cosmological models without singularity, 
Phys.\ Lett.\ B \textbf{91}, 99 (1980).

\bibitem{MukhanovChibisov1981}
V.\ F.\ Mukhanov and G.\ V.\ Chibisov,
Quantum fluctuations and a nonsingular universe,
JETP Lett.\ \textbf{33}, 532 (1981).

\bibitem{Starobinsky1982}
A.\ A.\ Starobinsky,
Dynamics of phase transition in the new inflationary universe 
scenario and generation of perturbations, 
Phys.\ Lett.\ B \textbf{117}, 175 (1982).

\bibitem{Starobinsky1983}
A.\ A.\ Starobinsky,
The Perturbation Spectrum Evolving from a Nonsingular Initially 
De-Sitter Cosmology and the Microwave Background Anisotropy,
Sov.\ Astron.\ Lett.\ \textbf{9}, 302 (1983).

\bibitem{Vilenkin1985}
A.\ Vilenkin,
Classical and quantum cosmology of the Starobinsky inflationary 
model,
Phys.\ Rev.\ D \textbf{32}, 2511 (1985). 

\bibitem{Linde2025}
A.\ Linde,
Alexei Starobinsky and Modern Cosmology,
2509.01675 [hep-th].

\bibitem{Capozziello2008}
S.\ Capozziello and M.\ Francaviglia,
Extended Theories of Gravity and their Cosmological and Astrophysical 
Applications, 
Gen.\ Rel.\ Grav.\ \textbf{40}, 357 (2008); 
arXiv:0706.1146 [astro-ph].

\bibitem{Capozziello2011}
S.\ Capozziello and M.\ De Laurentis, 
Extended Theories of Gravity,
Phys.\ Rept.\ \textbf{509}, 167 (2011); 
arXiv:1108.6266 [gr-qc].

\bibitem{Nojiri2007}
S.\ Nojiri and S.\ D.\ Odintsov, 
Introduction to modified gravity and gravitational alternative for 
dark energy,
Int.\ J.\ Geom.\ Meth.\ Mod.\ Phys.\ \textbf{4}, 115 (2007); 
arXiv:hep-th/0601213.

\bibitem{Nojiri2011}
S.\ Nojiri and S.\ D.\ Odintsov, 
Unified cosmic history in modified gravity: from $f(R)$ theory to 
Lorentz non-invariant models,
Phys.\ Rept.\ \textbf{505}, 59 (2011); 
arXiv:1011.0544 [gr-qc].

\bibitem{Nojiri2017}
S.\ Nojiri, S.\ D.\ Odintsov and V.\ K.\ Oikonomou,
Modified gravity theories on a nutshell: Inflation, bounce 
and late-time evolution,
Phys.\ Rept.\ \textbf{692}, 1 (2017); 
arXiv:1705.11098 [gr-qc].

\bibitem{DeFelice2010}
A.\ De Felice and S.\ Tsujikawa, 
$f(R)$ Theories,
Living Rev.\ Rel.\ \textbf{13}, 3 (2010); 
arXiv:1002.4928 [gr-qc].

\bibitem{Sotiriou2010}
T.\ P.\ Sotiriou and V.\ Faraoni, 
$f(R)$ Theories of Gravity,
Rev.\ Mod.\ Phys.\ \textbf{82}, 451 (2010); 
arXiv:0805.1726 [gr-qc].

\bibitem{PlanckInflation2018}
Y.\ Akrami \textit{et al.} [Planck Collaboration],
Planck 2018 results. X. Constraints on inflation,
Astron.\ Astrophys.\ \textbf{641}, A10 (2020);
arXiv:1807.06211 [astro-ph.CO].

\bibitem{BICEPKeck2021}
P.\ A.\ R.\ Ade \textit{et al.} [BICEP/Keck Collaboration],
Improved Constraints on Primordial Gravitational Waves using 
Planck, WMAP, and BICEP/Keck Observations through the 
2018 Observing Season,
Phys.\ Rev.\ Lett.\ \textbf{127}, 151301 (2021); 
arXiv:2110.00483 [astro-ph.CO].

\bibitem{DESI:2024}
A.\ G.\ Adame \textit{et al.} [DESI Collaboration],
DESI 2024 VI: Cosmological Constraints from the Measurements 
of Baryon Acoustic Oscillations, 
JCAP \textbf{02}, 021 (2025);
arXiv:2404.03002 [astro-ph.CO].

\bibitem{DESI:2025}
A.\ G.\ Adame \textit{et al.} [DESI Collaboration],
DESI 2024 VII: Cosmological Constraints from the Full-Shape 
Modeling of Clustering Measurements,
JCAP \textbf{07}, 028 (2025); 
arXiv:2411.12022 [astro-ph.CO].

\bibitem{ACT:2025}
T.\ Louis \textit{et al.} [ACT Collaboration],
The Atacama Cosmology Telescope: DR6 Power Spectra, Likelihoods 
and $\Lambda$CDM Parameters, 
arXiv:2503.14452 [astro-ph.CO].

\bibitem{ACT:2025-2}
E.\ Calabrese \textit{et al.} [ACT Collaboration],
The Atacama Cosmology Telescope: DR6 Constraints on Extended 
Cosmological Models, 
JCAP \textbf{11}, 063 (2025);
arXiv:2503.14454 [astro-ph.CO].

\bibitem{Ferreira2025}
E.\ G.\ M.\ Ferreira, E.\ McDonough, L.\ Balkenhol, R.\ Kallosh, 
L.\ Knox, and A.\ Linde,
BAO-CMB tension and implications for inflation,
Phys.\ Rev.\ D \textbf{113}, 043524 (2026);
arXiv:2507.12459 [astro-ph.CO].

\bibitem{McDonough2025}
E.\ McDonough and E.\ G.\ M.\ Ferreira,
The spectrum of $n_s$ constraints from DESI and CMB data,
arXiv:2512.05108 [astro-ph.CO].

\bibitem{HSQRT}
A.\ Galiautdinov,
Globally stable, ghost-free hyperbolic square-root deformation 
of the Starobinsky model,
arXiv:2603.09944 [gr-qc].

\bibitem{Kallosh2013AlphaAttractors}
R.\ Kallosh and A.\ Linde,
Universality Class in Conformal Inflation,
JCAP \textbf{07}, 002 (2013);
arXiv:1306.5220 [hep-th].

\bibitem{Kallosh2013AlphaAttractorsReview}
R.\ Kallosh, A.\ Linde, and D.\ Roest,
Superconformal inflationary $\alpha$-attractors,
JHEP \textbf{11}, 198 (2013);
arXiv:1311.0472 [hep-th].

\bibitem{QuantumGravity1}
I.\ L.\ Buchbinder, S.\ D.\ Odintsov, and I.\ L.\ Shapiro,
\textit{Effective Action in Quantum Gravity}
(IOP Publishing, Bristol, 1992).

\bibitem{QuantumGravity2}
I.\ L.\ Buchbinder and I.\ L.\ Shapiro,
\textit{Introduction to Quantum Field Theory with Applications 
to Quantum Gravity}
(Oxford University Press, 2021).

\bibitem{Barvinsky2023a}
A.\ O.\ Barvinsky\ and W.\ Wachowski,
Notes on conformal anomaly, nonlocal effective action, and 
the metamorphosis of the running scale,
Phys.\ Rev.\ D \textbf{108}, 045014 (2023); 
arXiv:2306.03780 [hep-th].

\bibitem{Barvinsky2023}
A.\ O.\ Barvinsky, G.\ H.\ S.\ Camargo, A.\ E.\ Kalugin, 
N.\ Ohta, and I.\ L.\ Shapiro,
Local term in the anomaly-induced action of Weyl quantum gravity,
Phys.\ Rev.\ D \textbf{108}, 086018 (2023).

\bibitem{Cognola2005}
G.\ Cognola, E.\ Elizalde, S.\ Nojiri, S.\ D.\ Odintsov and S.\ Zerbini,
One-loop $f(R)$ gravity in de Sitter universe,
JCAP \textbf{02}, 010 (2005); 
arXiv:0501096 [hep-th].

\bibitem{Mirzabekian1996}
A.\ G.\ Mirzabekian, G.\ A.\ Vilkovisky, V.\ V.\ Zhytnikov,
Partial summation of the nonlocal expansion for the 
gravitational effective action in 4 dimensions,
Phys.\ Lett.\ B \textbf{369}, 215 (1996).

\bibitem{Bonanno2002RGInflation}
A.\ Bonanno and M.\ Reuter,
Cosmology of the Planck era from a renormalization group for 
quantum gravity,
Phys.\ Rev.\ D \textbf{65}, 043508 (2002);
arXiv:hep-th/0106133.

\bibitem{Reuter2012AsymptoticSafety}
M.\ Reuter and F.\ Saueressig,
Quantum Einstein Gravity,
New J.\ Phys.\ \textbf{14}, 055022 (2012);
arXiv:1202.2274 [hep-th].

\bibitem{Alexandre2025RGFlowACT}
J.\ Alexandre, L.\ Heurtier, and S.\ Pla,
Exact Renormalisation Group Evolution of the Inflation Dynamics: 
Reconciling $\alpha$-Attractors with ACT,
arXiv:2511.05296 [hep-ph].

\bibitem{Baker1996}
G.\ A.\ Baker and P.\ Graves-Morris,
\textit{Pad\'{e} Approximants},
2nd ed. (Cambridge University Press, Cambridge, 1996).

\bibitem{Kallosh2025Status}
R.\ Kallosh and A.\ Linde,
On the present status of inflationary cosmology,
Gen.\ Rel.\ Grav.\ \textbf{57}, 135 (2025);
arXiv:2505.13646 [hep-th].

\bibitem{Martin2014}
J.\ Martin, C.\ Ringeval, and V.\ Vennin,
Encyclop{\ae}dia Inflationaris,
Phys.\ Dark Univ.\ \textbf{5-6}, 75 (2014); 
Phys.\ Dark Univ.\ \textbf{46}, 101653 (2024); 
arXiv:1303.3787 [astro-ph.CO].

\bibitem{Roest2014}
D.\ Roest, 
Universality classes of inflation, 
JCAP \textbf{01}, 007 (2014); 
arXiv:1309.1285 [hep-th].

\bibitem{GarciaBellido2014}
J.\ Garcia-Bellido and D.\ Roest,
Large-$N$ running of the spectral index of inflation,
Phys.\ Rev.\ D \textbf{89}, 103527 (2014); 
arXiv:1402.2059 [astro-ph.CO].

\bibitem{Kallosh2022Polynomial}
R.\ Kallosh and A.\ Linde, 
Polynomial $\alpha$-attractors,
JCAP \textbf{04}, 017 (2022);
arXiv:2202.06492 [astro-ph.CO].

\bibitem{Kallosh2025Singular}
R.\ Kallosh and A.\ Linde,
Singular $\alpha$-attractors,
arXiv:2512.02969 [hep-th].

\bibitem{Gialamas2025ACTive}
I.\ D.\ Gialamas, T.\ Katsoulas, and K.\ Tamvakis,
Keeping the relation between the Starobinsky model 
and no-scale supergravity ACTive,
JCAP \textbf{09}, 060 (2025);
arXiv:2505.03608 [gr-qc].

\bibitem{Kallosh2025ACT}
R.\ Kallosh, A.\ Linde, and D.\ Roest,
ACT, SPT, and chaotic inflation,
arXiv:2503.21030 [hep-th].

\bibitem{Gialamas2026Weyl}
I.\ D.\ Gialamas,
Reheating in geometric Weyl-invariant Einstein-Cartan gravity,
arXiv:2602.00317 [gr-qc].

\bibitem{Sorokin2022}
D.\ P.\ Sorokin, 
Introductory Notes on Non-linear Electrodynamics and its Applications, 
Fortschr.\ Phys.\ \textbf{70}, 2200092 (2022); 
arXiv:2112.12118 [hep-th].

\bibitem{Yuennan2025}
J.\ Yuennan, F.\ Atamurotov, and P.\ Channuie,
ACT Constraints on Marginally Deformed Starobinsky Inflation,
Phys.\ Lett.\ B \textbf{872}, 140065 (2026);
arXiv:2509.23329 [gr-qc].

\bibitem{Addazi2025}
A.\ Addazi, Y.\ Aldabergenov, S.\ V.\ Ketov,
Curvature corrections to Starobinsky inflation can explain the ACT results,
Phys.\ Lett.\ B \textbf{869}, 139883 (2025);
arXiv:2505.10305 [gr-qc].

\bibitem{Bezerra-Sobrinho2025}
J.\ Bezerra-Sobrinho, L.\ G.\ Medeiros,
Starobinsky Inflation and the Latest CMB Data: A Subtle Tension?,
arXiv:2511.06640 [astro-ph.CO].

\bibitem{Odintsov2026}
S.\ D.\ Odintsov, V.\ K.\ Oikonomou, and G.\ S.\ Sharov,
Viable $f(R)$ scenarios unifying inflation with realistic dynamical dark energy,
JHEA \textbf{52}, 100579 (2026);
arXiv:2601.06949 [gr-qc].

\bibitem{Kamenshchik2016}
A.\ Yu.\ Kamenshchik, E.\ O.\ Pozdeeva, S.\ Yu.\ Vernov, 
A.\ Tronconi, and G.\ Venturi,
Transformations between Jordan and Einstein frames: Bounces, 
antigravity, and crossing singularities,
Phys.\ Rev.\ D \textbf{94}, 063510 (2016); 
arXiv:1602.07192 [gr-qc].

\bibitem{Hamilton1982}
R.\ S.\ Hamilton, 
Three-manifolds with positive Ricci curvature, 
J.\ Diff.\ Geom.\ \textbf{17}, 255 (1982).

\bibitem{Perelman2002}
G.\ Perelman, 
The entropy formula for the Ricci flow and 
its geometric applications, 
arXiv:math/0211159 [math.DG].

\bibitem{Friedan1980}
D.\ Friedan, 
Nonlinear Models in $2+\epsilon$ Dimensions, 
Phys.\ Rev.\ Lett.\ \textbf{45}, 1057 (1980).

\bibitem{Friedan1985}
D.\ Friedan, 
Nonlinear Models in $2+\epsilon$ Dimensions, 
Ann.\ Phys.\ \textbf{163}, 318 (1985).

\end{thebibliography}
\end{document}